\documentclass[%
 reprint,
 superscriptaddress,
 amsmath,amssymb,
 aps,https://www.overleaf.com/project/5d5ae32d4618dd37df01b4f9
 pra,
floatfix,
]{revtex4-1}
\usepackage{amsmath}
\usepackage{lipsum}

\usepackage{multirow}
\usepackage{graphicx}
\usepackage{dcolumn}
\usepackage{bm}
\usepackage[colorlinks,linkcolor=blue,citecolor = blue,urlcolor=blue]{hyperref}
\usepackage{booktabs}
\usepackage{float}
\usepackage{csquotes}
\usepackage{abraces}

\usepackage{commath}
\usepackage{ulem}
\usepackage{braket}
\usepackage{color}
\usepackage{cleveref}
\usepackage{array}
\usepackage{caption}

\usepackage{physics}

\begin{document}

\title{Asynchronous Multi-photon Interference for Quantum Networks}

\author{Baghdasar Baghdasaryan}
\email{baghdasar.baghdasaryan@uni-jena.de}
\affiliation{Friedrich Schiller University Jena, Abbe Center of Photonics, Institute of Applied Physics, 07745 Jena, Germany}

\author{Karen Lozano-Méndez}
\affiliation{Fraunhofer Institute for Applied Optics and Precision Engineering IOF, 07745 Jena, Germany}

\author{Markus Leipe}
\affiliation{Friedrich Schiller University Jena, Abbe Center of Photonics, Institute of Applied Physics, 07745 Jena, Germany}
\affiliation{Fraunhofer Institute for Applied Optics and Precision Engineering IOF, 07745 Jena, Germany}

\author{Meritxell Cabrejo-Ponce}
\affiliation{Friedrich Schiller University Jena, Abbe Center of Photonics, Institute of Applied Physics, 07745 Jena, Germany}
\affiliation{Fraunhofer Institute for Applied Optics and Precision Engineering IOF, 07745 Jena, Germany}

\author{Sabine Häussler}
\affiliation{Friedrich Schiller University Jena, Abbe Center of Photonics, Institute of Applied Physics, 07745 Jena, Germany}
\affiliation{Fraunhofer Institute for Applied Optics and Precision Engineering IOF, 07745 Jena, Germany}

\author{Kaushik Joarder}
\affiliation{Friedrich Schiller University Jena, Abbe Center of Photonics, Institute of Applied Physics, 07745 Jena, Germany}

\author{Tim Gühring}
\affiliation{Fraunhofer Institute for Applied Optics and Precision Engineering IOF, 07745 Jena, Germany}

\author{Stephan Fritzsche}
\affiliation{Friedrich Schiller University Jena, Institute for Theoretical Physics, 07743 Jena, Germany}
\affiliation{Helmholtz Institute Jena, 07743 Jena, Germany}

\author{Thorsten A. Goebel}
\affiliation{Friedrich Schiller University Jena, Abbe Center of Photonics, Institute of Applied Physics, 07745 Jena, Germany}
\affiliation{Fraunhofer Institute for Applied Optics and Precision Engineering IOF, 07745 Jena, Germany}

\author{Ria G. Krämer}
\affiliation{Friedrich Schiller University Jena, Abbe Center of Photonics, Institute of Applied Physics, 07745 Jena, Germany}

\author{Stefan Nolte}
\affiliation{Friedrich Schiller University Jena, Abbe Center of Photonics, Institute of Applied Physics, 07745 Jena, Germany}
\affiliation{Fraunhofer Institute for Applied Optics and Precision Engineering IOF, 07745 Jena, Germany}

\author{Carlos Andres Melo Luna}
\affiliation{Fraunhofer Institute for Applied Optics and Precision Engineering IOF, 07745 Jena, Germany}

\author{Yoshiaki Tsujimoto}
\affiliation{National Institute of Information and Communications Technology (NICT), 184-8795 Tokyo, Japan}

\author{Fabian Steinlechner}
\email{fabian.steinlechner@uni-jena.de}
\affiliation{Friedrich Schiller University Jena, Abbe Center of Photonics, Institute of Applied Physics, 07745 Jena, Germany}
\affiliation{Fraunhofer Institute for Applied Optics and Precision Engineering IOF, 07745 Jena, Germany}

\date{\today}

\begin{abstract}
Advanced quantum communication protocols require high-visibility quantum interference between photons generated at distant nodes, which places stringent demands on optical synchronization. Conventionally, synchronization of optical wave packets relies on pulsed sources and precise optical path stabilization. An alternative approach employs continuous-wave (CW) photon-pair sources, where temporal indistinguishability is enforced by post-selecting detection events within a coincidence window $\tau_w$ shorter than the photon coherence time $T_c$. Despite its conceptual simplicity, the quantitative relation between relevant time scales, achievable interference visibility, and usable multi-photon rates has remained unclear. Here, we develop in detail and experimentally validate a theoretical framework that quantitatively describes time-resolved multi-photon interference in the CW regime. We explicitly incorporate detector timing jitter, photon coherence time, and temporal post-selection. The model is verified using four-photon Hong-Ou-Mandel interference measurements. Based on this validated framework, we determine the coincidence window that maximizes usable four-photon rates for a target visibility. Finally, we compare CW and pulsed SPDC sources under equivalent indistinguishability constraints and show that CW operation can achieve comparable rates while relaxing optical synchronization requirements.
\end{abstract}

\pacs{Valid PACS appear here}
\maketitle
\section{Introduction}
In recent years, interest in photonic quantum technologies and quantum networks has grown rapidly. This progress drives demand for robust multi-photon quantum interference, an essential building block for measurement-device-independent quantum key distribution \cite{Wang2021}, boson sampling \cite{PhysRevA.101.063821}, photonic quantum repeaters \cite{Azuma2015}, quantum teleportation \cite{Ren2017}, and entanglement swapping \cite{PhysRevLett.71.4287}. 

\begin{figure*}[t!]
 \center
\includegraphics[width=0.9\textwidth]{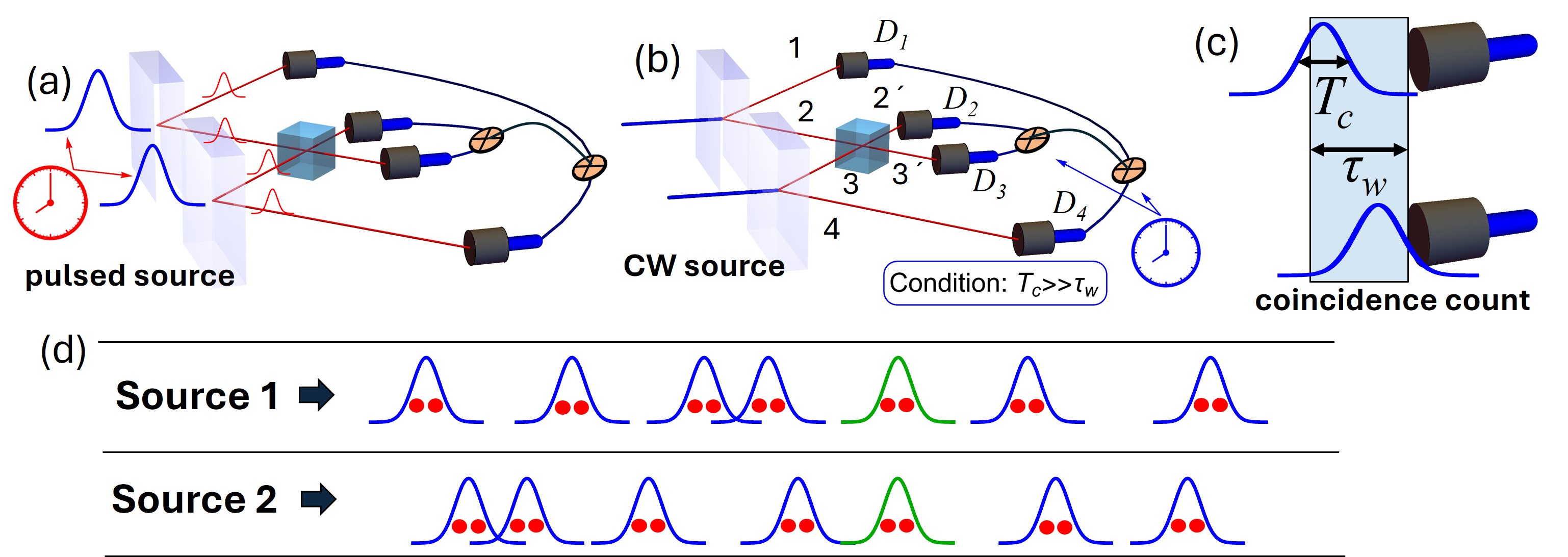}
\caption{(a) Realization of temporal indistinguishability in entanglement swapping via time-synchronized pulsed photon sources. (b) Entanglement swapping with continuous-pumped photon pair sources. Temporally indistinguishable photons can be post-selected with single photon detectors if the coherence time of photons $T_c$ exceeds the coincidence window $\tau_w$ of detection events. (c) Detection of two photon wave packets with given coincidence window $\tau_w$. Since the coincidence window is larger than the coherence time $T_c$, photon pairs with large temporal distances can be inaccurately counted as a simultaneous coincidence. (d) Continuous emission of photon pairs from two CW sources, where among all pairs, the temporally indistinguishable photons (green) can be post-selected.}
\label{fig1}    
\end{figure*}

High-visibility quantum interference requires the interfering photons to be indistinguishable in all degrees of freedom \cite{Qian2023,PhysRevLett.133.233601}. The maximum achievable visibility is bounded by the single-photon purity that is commonly engineered by tailoring the phase-matching condition and group velocities in nonlinear media \cite{PhysRevA.64.063815,Garay-Palmett:07,Branczyk:11,BenDixon:13,PhysRevLett.100.133601,PhysRevA.93.013801,Tambasco:16,Graffitti:18,Pickston:21,PhysRevA.108.023718} or by applying spectral filtering \cite{PhysRevA.96.053842}. Beyond spectral considerations, temporal indistinguishability must also be enforced. In pulsed implementations based on spontaneous parametric down-conversion (SPDC) \cite{PhysRevLett.80.3891,PhysRevLett.88.017903,PhysRevLett.96.110501,Jin2015}, pair creation is confined to short pump pulses [Fig.~\ref{fig1}(a)]. The interference visibility in the pulsed regime depends on the single-photon wave-packet duration and the precise synchronization of both the pump pulses and optical paths\cite{PhysRevLett.59.2044,doi:10.1139/cjp-2023-0312}. While this approach is well-suited to controlled laboratory settings, extending it to long-distance or distributed quantum networks requires active stabilization of emission times and path lengths across remote nodes, which are major challenges towards deployment in real network infrastructure \cite{Sun2016}.

An alternative approach, first proposed in \cite{PhysRevLett.71.4287} and later demonstrated experimentally \cite{Halder2007, Halder_2008,Tsujimoto2018}, employs photon sources pumped by continuous-wave (CW) lasers. In the CW regime, photon pairs are emitted randomly in time at a steady rate [Fig.~\ref{fig1}(b)]. Temporal indistinguishability is then enforced by post-selecting detection events within a narrow coincidence window $\tau_w$ satisfying $\tau_w \ll T_c$ \cite{Halder_2008,PhysRevA.82.043826}, where $T_c$ denotes the coherence time of interfering photons. Physically, this corresponds to single-temporal-mode detection, in which the coincidence window acts as a temporal projector that selects indistinguishable emission events [Fig.~\ref{fig1}(c,d)].

Importantly, while this approach removes the requirement for optical synchronization of photon emission times, it does not eliminate the need for classical clock synchronization for the detection of correlated events. Instead, asynchronous CW operation shifts the synchronization task from optical emission-time control to electronic clock synchronization, which is generally less demanding and easier to scale \cite{PhysRevApplied.19.054082,Spiess_2024}.

Despite the availability of highly coherent quantum light sources \cite{Samara:19} and ultra-fast detectors \cite{Natarajan_2012} capable of accessing the single-temporal-mode regime \cite{Halder2007}, a quantitative design framework for the CW approach is still lacking. In particular, it remains unclear how strictly the condition $\tau_w \ll T_c$ needs to be satisfied in order to meet application-specific visibility thresholds--e.g., $\geq50\%$ for non-classical two-photon interference \cite{Moschandreou:18}, $\geq71\%$ for Bell inequality violation \cite{PhysRevA.78.032112}, or $\geq 95\%$ for cluster-state growth \cite{PRXQuantum.6.020304}. Moreover, the impact of finite detector timing jitter $j$ on indistinguishability has not been analyzed quantitatively. These questions are central to assessing the scalability of asynchronous quantum networks driven by CW sources.

Here, we develop and experimentally validate a theoretical model that quantitatively describes photon indistinguishability in time-resolved Hong-Ou-Mandel (HOM) interference experiments in the CW SPDC regime \cite{Legero2003,PhysRevLett.93.070503,LEGERO2006253,Tsujimoto:17}. We show that, in the low-noise limit, the visibility is predominantly governed by the ratio $T_c/\tau_w$. We verify this scaling experimentally using sources with different coherence times, without any post-hoc fitting. Building on this validated model, we analyze the attainable four-photon rates in a simple quantum-network scenario based on CW SPDC sources. We demonstrate that, for a given target visibility and fixed timing jitter $j$, an optimal coincidence window exists that maximizes the usable photon rate. Finally, we identify the parameter regime in which CW operation becomes advantageous for scalable quantum networks and compare its performance to pulsed SPDC sources under equivalent indistinguishability constraints.%
\section{Multi-photon interference in the CW regime}
We analyze a HOM-type multi-photon interference in the context of entanglement swapping, which is an essential protocol for quantum repeaters in global quantum networks \cite{PhysRevLett.81.5932,Duan2001,PhysRevLett.90.207901,PhysRevLett.101.080403,Liao2017,doi:10.1126/science.aan3211}. A schematic of entanglement swapping in the CW regime is shown in Fig. \ref{fig1} (b). Entanglement can be encoded in time \cite{Halder2007,PhysRevA.95.032306,Sun:17,Samara_2021}, frequency \cite{PhysRevLett.128.063602}, polarization \cite{PhysRevA.77.022312,Tsujimoto2018,PhysRevLett.123.160502}, or orbital angular momentum \cite{Zhang2017}. Our subsequent analysis focuses exclusively on temporal indistinguishability, independent of the specific encoding degree of freedom.

\begin{figure}
\includegraphics[width=.5\textwidth]{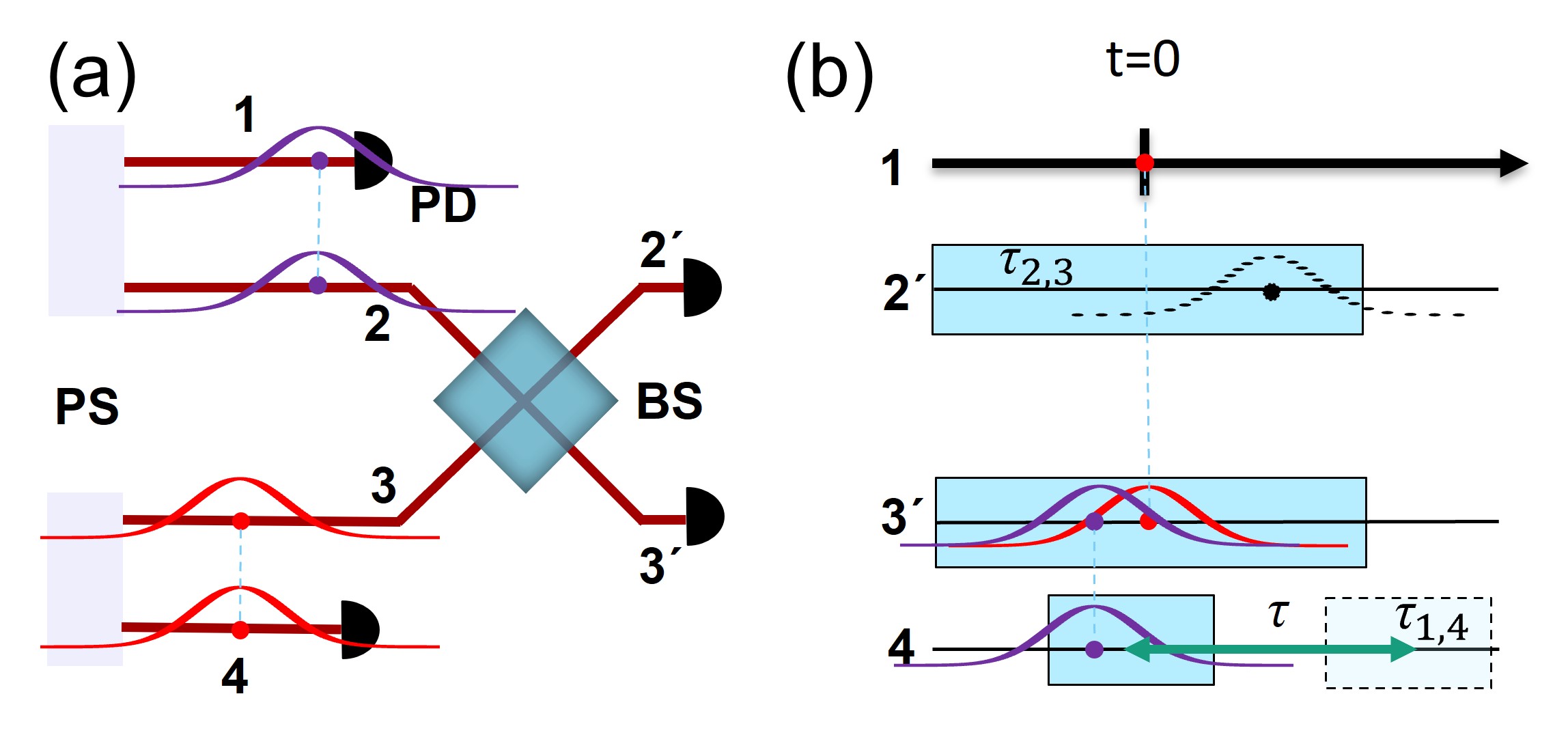}
\caption{(a) We assume the outer photons are detected first. The coincidence window $\tau_{1,4}$ sets the maximum allowed time difference between the detection events in the spatial modes $1$ and $4$, if $j<\tau_{1,4}$ is ensured. Because of the strong time correlation within each pair, this $\tau_{1,4}$ window predetermines the temporal separation—and thus the indistinguishability-of the photons in modes $2$ and $3$. The coincidence windows $\tau_{2,3}$ in spatial modes $2^{\prime}$ and $3^{\prime}$ does not affect this indistinguishability; it determines the probability of successfully capturing the wave packets of these photons. (b) The illustration of the coincidence windows relative to the trigger photon in the spatial mode $1$. This example illustrates that the photons exit into the same spatial mode $3^{\prime}$, leaving $2^{\prime}$ empty. A noise photon (dashed wave packet) can then occupy $2^{\prime}$ and produce a false fourfold. The dashed window in mode 4 indicates the $\tau$-dependent shift of the $\tau_{1,4}$ coincidence window.  PD: photon detector; PS: photon-pair source.}\label{window}
\end{figure}

Temporal indistinguishability is assessed through four-photon detection events in the setup shown in Fig.~\ref{window} (a). If photons in spatial modes $2$ and $3$ are indistinguishable and arrive simultaneously at the beam splitter (BS), they exit through the same output mode ($2^{\prime}$ or $3^{\prime}$), which gives rise to the well-known HOM dip \cite{PhysRevLett.59.2044,Bouchard_2021,doi:10.1139/cjp-2023-0312}. In a four-photon setting, this interference directly mirrors the Bell-state measurement required for entanglement swapping.

To describe this process theoretically, we begin with the biphoton states generated by two independent CW sources and then sequentially propagate the states through optical elements and apply detection operators. The results derived below apply to any photon-pair source operating in the CW regime. The biphoton state produced by a monochromatic CW source can be written as \cite{PhysRevA.106.063711,PhysRevA.109.023534}
 \begin{equation}
    \ket{\Psi}_{i,k} =N \int  d\Omega\,  \Phi_{i,k}(\Omega) \,\ket{\Omega}_i\ket{-\Omega}_k\label{SPDCstate},
\end{equation}
where $i$ and $k$ denote the propagation modes, $N$ is the normalization factor, $\Omega$ is the frequency deviation from the center frequency $\omega_0$ ($\omega=\omega_0\pm \Omega$). The biphoton mode function $\Phi_{i,k}(\Omega)$ describes the spectral structure of the source arising from phase matching and dispersion.

For broadband sources, spectral filtering is required to extend the photon coherence time such that $\tau_w \ll T_c$. Narrowband filters are modeled as 
\begin{equation*}
  \hat{\mathcal{F}}_i=\int\,d\Omega\, F_i(\Omega)\, \ket{\Omega}_i\bra{\Omega}_i,
\end{equation*}
where $F_i(\Omega)$ denotes the filter transmission function. Applying this operator to the biphoton state yields
 \begin{equation}
    \ket{\Psi}_{i,k}\propto \int  d\Omega\,J_{i,k}(\Omega) \,\ket{\Omega}_i\ket{-\Omega}_k\label{Fstate},
\end{equation}
where we introduced the following notation for the joint spectral amplitude:
\begin{equation*}
    J_{i,k}(\Omega)=F_i(\Omega)\,F_k(-\Omega)\,  \Phi_{i,k}(\Omega).
\end{equation*}
Note that the quantum state $\ket{\Psi}_{i,k}$ is primarily determined by the filter response if the filter bandwidth $F_i(\Omega)$ is much narrower than the intrinsic biphoton spectrum $\Phi_{i,k}(\Omega)$.

For two independent sources, the joint state of two photon pairs is%
 \begin{align*}
  \nonumber   \ket{\Psi}_{1,2}\ket{\Psi}_{3,4} &\propto \iint  d\Omega\,d\Omega^{\prime}\, J_{1,2}(\Omega)\,J_{3,4}(\Omega^{\prime}) \\& \:\ket{\Omega}_1\ket{-\Omega}_2\ket{-\Omega^{\prime}}_3\ket{\Omega^{\prime}}_4.
\end{align*}
Next, we apply a $50/50$ BS on the photons in modes $2$ and $3$ and retain only the terms in which the photons exit in different spatial modes. The resulting normalized state is given by
 \begin{align}
 \nonumber   \ket{\Psi}_{1,2^{\prime},3^{\prime},4}&=N^{\prime} \iint  d\Omega\,d\Omega^{\prime}\, J_{1,2}(\Omega)\,J_{3,4}(\Omega^{\prime})  \\& \:\ket{\Omega}_1\ket{\Omega^{\prime}}_4\,\bigl(\ket{-\Omega}_{2^{\prime}}\ket{-\Omega^{\prime}}_{3^{\prime}}-\ket{-\Omega^{\prime}}_{2^{\prime}}\ket{-\Omega}_{3^{\prime}}\bigr),\label{composite}
\end{align}
where $N^{\prime}$ is the new normalization factor.

Photon detection at time $t_0$ with perfect time resolution would be described by the projector $\ket{t_0}\bra{t_0}$. In practice, finite detector timing resolution must be included through the detector response function $g_i(t)$ \cite{PhysRevA.98.013833}. We model the resulting measurement as the POVM element
\begin{equation}
    \hat{P_i}(t_0)=\int\,dt\, g_i(t) \ket{t+t_0}_i\bra{t+t_0}_i.\label{detector}
\end{equation}
Here, the detector efficiency is assumed to be unity and frequency independent. The response function $g_i(t)$ accounts for timing uncertainty arising from detector jitter $j$, which we model as a Gaussian with full width at half maximum $j$.

To express both the state from Eq.~\eqref{composite} and the detection operators in the same basis, we transform the detection operator into the frequency domain,
\begin{equation}
    \hat{P}_i(t_0)=\iint\,d\Omega_1\, d\Omega_2\, g_{F,i}(\Omega_1-\Omega_2)\, e^{it_0(\Omega_1-\Omega_2)} \ket{\Omega_1}_{i}\bra{\Omega_2}_{i},\label{det}
\end{equation}
where $a^{\dagger}(t)=\frac{1}{\sqrt{2\pi}}\int d \Omega\, a^{\dagger}(\Omega) \,e^{i\Omega t}$, and  $g_{F,i}(\Omega_1-\Omega_2)$ is proportional to the Fourier transform of the detector response function 
\begin{equation*}
    g_{F,i}(\Omega_1-\Omega_2)=\frac{1}{2\pi}\int\,dt\, g_i(t)\, e^{it(\Omega_1-\Omega_2)}.
\end{equation*}
We now have all the components to describe the detection process. The probability that photons are detected at times $t_1,t_2,t_3,t_4$ is then
\begin{equation*}
  p(t_1,t_2,t_3,t_4)=  \mathrm{Tr}\bigl( \hat{P}_1(t_1)\,\hat{P}_{2^{\prime}}(t_2)\,\hat{P}_{3^{\prime}}(t_3)\,\hat{P}_4(t_4)\,\rho\bigr),
\end{equation*}
where $\rho= \ket{\Psi}_{1,2^{\prime},3^{\prime},4}\bra{ \Psi}_{1,2^{\prime},3^{\prime},4}$ is the density operator of the composite system Eq.~\eqref{composite}.

Next, we must include the finite coincidence windows used in the experiment. Photons in modes $1$ and $4$ serve as heralding (outer) photons, while photons in modes $2$ and $3$ interfere at the BS. The coincidence window between detections in modes $1$ and $4$ is denoted $\tau_{1,4}$ and defines the maximum allowed time difference between these detection events (assuming $j<\tau_{1,4}$). A smaller $\tau_{1,4}$ therefore enforces tighter temporal synchronization.
Due to strong time correlations within each photon pair, the coincidence window $\tau_{1,4}$ predetermines the temporal separation--and thus the indistinguishability--of the photons interfering in modes $2$ and $3$. In contrast, the coincidence windows $\tau_{2,3}$ only determines the probability of capturing the interfering wave packets and should therefore be chosen larger than the coherence time, $\tau_{2,3}>T_c$. In Fig. \ref{window} (b), the coincidence windows are presented by blue rectangles, where the photons in the spatial mode $1$ are used for triggering.

To calculate the four-photon coincidence probability, we integrate over all detection times within the coincidence windows. All optical paths are assumed to be equal. The detection times of photons in modes $2^{\prime}$ and $3^{\prime}$ are set the same and are integrated around a reference time $t=0$ within $[-\tau_{2,3}/2,\tau_{2,3}/2]$. A variable arrival time $\tau$ is introduced for photons in the spatial mode 4 [see Fig. \ref{window} (b)], to test how the interference visibility changes when photons lose temporal overlap. For given $\tau_{1,4}$, the highest interference is expected at $\tau=0$ and no interference is expected at $\tau\gg T_c$. The integration range for $t_4$ is therefore $(\tau-\tau_{1,4}/2,\tau+\tau_{1,4}/2)$.

The integration over the coincidence windows for each value of $\tau$ yields:
\begin{align}
      \mathcal{P}(\tau)&= \int_{-\tau_{2,3}/2}^{\tau_{2,3}/2}\int_{-\tau_{2,3}/2}^{\tau_{2,3}/2}\nonumber\\&  \int_{\tau-{\tau_{1,4}/2}}^{\tau+{\tau_{1,4}/2}}
   \,dt_2\, dt_3\, dt_4 \,p(0,t_2,t_3,t_4).\label{HOM1} 
\end{align}

After applying the detection operators and evaluating the time integrals in Eq.~\eqref{HOM1}, we end up with the following expression for the four-photon coincidence probability:
\begin{align}
\nonumber\mathcal{P}(\tau) &= \iiiint d\Omega_1\,d\Omega_2\,d\Omega_3\,d\Omega_4\:{\tau_{1,4}}\,  \mathrm{sinc}\left[\frac{\tau_{1,4}}{2}(\Omega_4-\Omega_3)\right] \\\nonumber
&\times J_{1,2}(\Omega_1)\,J^*_{1,2}(\Omega_2)\,J_{3,4}(\Omega_3)\,J^*_{3,4}(\Omega_4) \\
&\times g_{F,1}(\Omega_2-\Omega_1)g_{F,4}(\Omega_4-\Omega_3)e^{i\tau(\Omega_4-\Omega_3)} \nonumber\\
&\nonumber\times\Bigl[ \phi_2(\Omega_1,\Omega_2)\,\phi_3(\Omega_3,\Omega_4)  + \phi_2(\Omega_3,\Omega_4)\,\phi_3(\Omega_1,\Omega_2) \\
& \quad -\phi_2(\Omega_3,\Omega_2)\,\phi_3(\Omega_1,\Omega_4) - \phi_2(\Omega_1,\Omega_4)\,\phi_3(\Omega_3,\Omega_2) \Bigr], \label{HOMmain}
\end{align}
where
%
\begin{equation*}
\phi_i(\Omega_1,\Omega_2)= \tau_{2,3}\,\mathrm{sinc}\left[\frac{\tau_{2,3}}{2}(\Omega_2-\Omega_1)\right] g_{F,i}(\Omega_2-\Omega_1).
\end{equation*}
Here, \textit{sinc} functions arise from the convolution integral over the rectangular coincidence windows.
The resulting expression for the four-photon coincidence probability $\mathcal{P}(\tau)$ depends on the delay $\tau$, the coincidence windows $\tau_{1,4}$ and $\tau_{2,3}$, the detector timing jitter $j$, and the spectral filter response function.
\begin{figure*}[t!]
 \center
\includegraphics[width=0.9\textwidth]{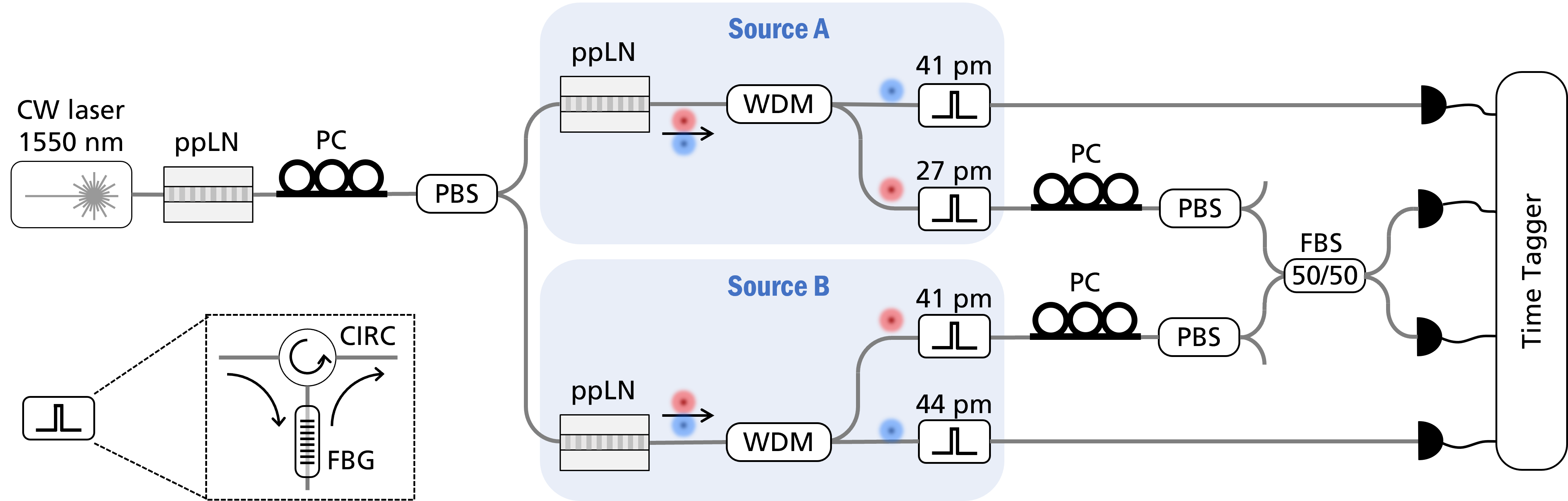}
\caption{Experimental setup for performing four-photon HOM interference. A CW telecom laser is up-converted and then split into two paths to pump two SPDC sources, A and B (blue boxes). The generated two-photon states are separated with a wavelength division multiplexer (WDM) according to their higher or lower frequencies, followed by tight filtering with a fiber Bragg grating, as described in the left inset. The signal photons (blue) are used to herald the interference, while the idler photons (red) are interfered at a polarization-maintaining fiber beam splitter (FBS). CIRC: circulator; PBS: polarization beam splitter; PC: polarization controller; ppLN: periodically-poled lithium niobate waveguide.}
\label{Expsetup}    
\end{figure*}
\section{Experimental setup}
\begin{figure}
\includegraphics[width=.5\textwidth]{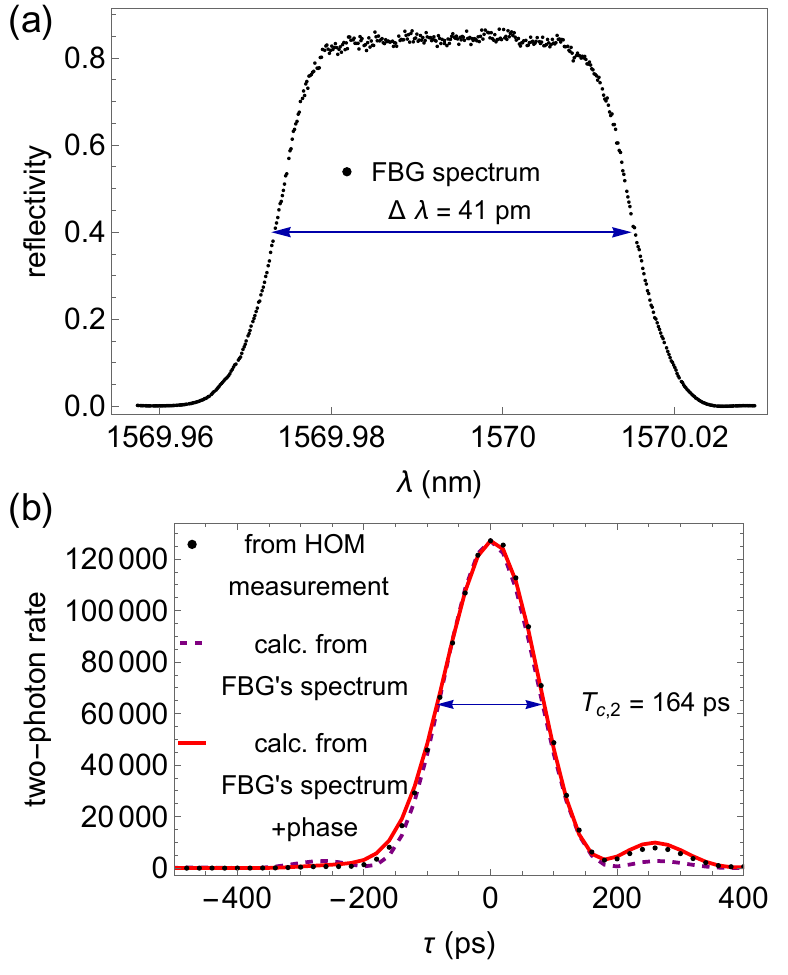}
\caption{(a) Normalized reflection spectrum of the FBG filter with a bandwidth of $\Delta\lambda=41$\,pm. The wavelength-dependent reflectance was determined by normalizing the reflected signal to the incident probe spectrum. (b) Two-photon coincidence counts for the source B (dots) compared with the calculated and normalized coherence function without (dashed purple curve) and with the phase model (solid red curve).}\label{filter}
\end{figure}
The experimental setup shown in Fig. \ref{Expsetup} consists of a $775$ nm pump laser obtained by second-harmonic generation (SHG) from a continuous wave laser at $1550$\,nm using a fiber-coupled, periodically poled lithium niobate (ppLN) waveguide. The pump laser is split using a fiber polarizing beam splitter (PBS) to drive SPDC emission in two separate ppLN waveguides. The waveguides generate correlated photons centered at $1550$\,nm via type-0 SPDC. Signal and idler photons are separated using wavelength-division multiplexers and spectrally filtered with fiber Bragg gratings (FBGs). The filters include femtosecond-inscribed in-house devices \cite{Kraemer2025} and commercial FBGs (Teraxion). The transmission spectra of the FBGs have a full width at half maximum (FWHM) of $27$\,pm and $41$\,pm for the idlers and $41$\,pm and $44$\,pm for signals. To erase polarization distinguishability, the photons are coupled into a fiber PBS, and their polarization is optimized using fiber polarization controllers. The idler photons are then interfered on a  50/50 BS. We note that placing the idler filters downstream of the 50/50 BS would suppress which-path information associated with different filter widths and is therefore the preferable configuration in high-visibility entanglement-swapping experiments. In the present implementation, the filters were positioned before the BS, whereby different filter bandwidths enabled benchmarking the observed experimental imperfections against the theoretical model.

All four-photons were detected using superconducting nanowire single photon detectors (SNSPD, Single Quantum). Detection events are recorded by a time-tagging unit (quTools) with an RMS timing jitter of $4.2$ ps. In the theoretical analysis, the overall temporal response is modeled as the convolution of the time-tagger and detector timing jitters.

\section{Results and Discussion}
\subsection{Temporal coherence function of biphoton state}
We first experimentally characterize the temporal coherence of the generated photon pairs. The temporal coherence of the biphoton wave packet is determined by the spectral amplitude and phase of the applied FBG filters. Figure \ref{filter} (a) shows the reflected power spectrum of $\Delta\lambda=41$\,pm FBG filter obtained by an optical spectrum analyzer. Since this measurement provides only the spectral magnitude, we reconstruct the spectral phase using a transfer-matrix model of the FBG \cite{618322}. A super-Gaussian refractive-index profile is assumed, and its parameters are optimized to reproduce the measured reflection spectrum. The spectral phase is then extracted from the resulting complex transfer function.

The temporal coherence function is obtained from Eq.~\eqref{Fstate} by applying the detection operators Eq.~\eqref{det} on signal and idler photons with relative time delay $\tau$:
\begin{align*}
   G_{i,k}(\tau)= \iint\,d\Omega_1\,&d\Omega_2\,J_{i,k}(\Omega_1)\,J_{i,k}^{\ast}(\Omega_2)\,e^{i(\Omega_1-\Omega_2)\tau}\\&\times g_{F,i}(\Omega_2-\Omega_1)\,g_{F,k}(\Omega_1-\Omega_2).
\end{align*}
This function represents the probability density of detecting an idler photon at time delay $\tau$ relative to the signal photon. 

Figure~\ref{filter}(b) compares the measured two-photon coincidences of source B with the calculated temporal coherence function. Including the reconstructed spectral phase (solid red curve) reproduces the experimentally observed side peak, whereas a phase-free model (dashed purple curve) fails to capture this feature. Both experiment and theory yield a coherence time of $T_{c,2}=164$\,ps defined as FWHM. The same procedure yields $T_{c,1}=244$\,ps for source A. Consistency is confirmed by computing the coherence function directly from Eq.~\eqref{HOMmain} by tracing over the arrival times of the remaining photons, i.e., assuming infinitely large coincidence windows.
\subsection{HOM measurement}
\begin{figure}
\includegraphics[width=.5\textwidth]{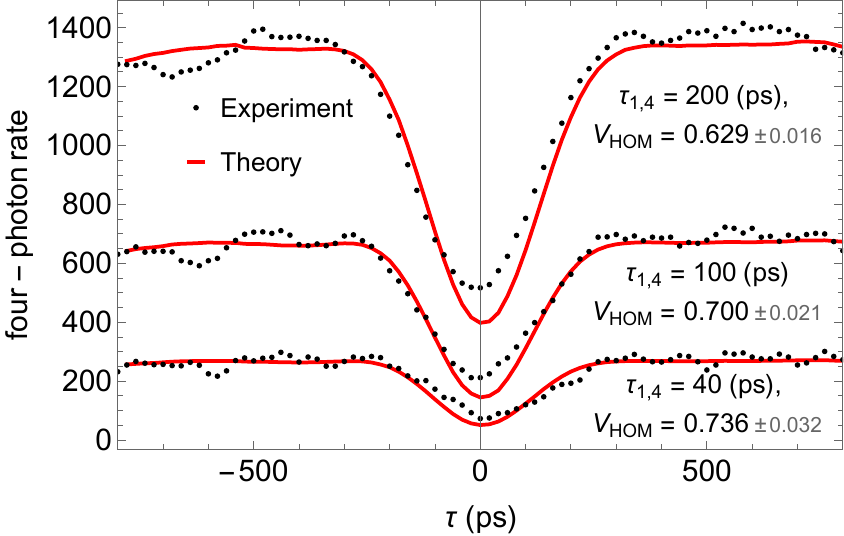}
\caption{Time-resolved HOM dip for different coincidence windows $\tau_{1,4}$ of the outer photons, where the coincidence window of the photons in modes $2^{\prime}$ and $3^{\prime}$ is set to $\tau_{2,3}= 2000$\,ps. For the theoretical calculation, we normalized Eq.~\eqref{HOMmain} according to experimental photon count rate for $\tau_{1,4}= 40$\,ps. The same normalization is used for the rest of the theoretical prediction. The total rate was measured over a $1.5$-hour period.}\label{HOMcurves}
\end{figure}

Figure~\ref{HOMcurves} shows the measured and predicted four-photon coincidence rate as a function of the detection delay in spatial mode~4 for several coincidence windows $\tau_{1,4}$. The coincidence windows between the trigger photon and the photons in modes $2^{\prime}$ and $3^{\prime}$ were chosen much larger than the photon coherence times, $\tau_{2,3}=2000$\,ps $\gg T_{c,1},T_{c,2}$, such that temporal post-selection is governed solely by $\tau_{1,4}$. As the detection delay increases, the two-photon wave packets become temporally distinguishable, and the coincidence rate rises, yielding the expected time-resolved HOM dip.

Solid curves in Fig.~\ref{HOMcurves} show the theoretical predictions obtained from Eq.~\eqref{HOMmain}. Detector timing jitters of $j=(17,13,11,16)$\,ps for modes $1$–$4$ were used directly without post-hoc fitting; the only free scale parameter is a global normalization factor.

The theoretical probability $\mathcal{P}(\tau)$ was normalized to the experimental plateau at $\tau\gg T_c$ for the data with $\tau_{1,4}=40$\,ps after correction for accidental four-fold events (detailed in Sec. \ref{accSec}) and subsequently applied to all theoretical curves. The model reproduces the measured HOM dips across all coincidence windows. The slightly elevated experimental offsets at larger $\tau_{1,4}$ are consistent with an increased contribution of accidental events.

All curves in Fig.~\ref{HOMcurves} exhibit a HOM dip, yet retain a nonzero coincidence rate at $\tau=0$ due to imperfect visibility arising from residual photon distinguishability due to a finite coincidence window, as well as accidental counts. To quantify the degree of indistinguishability of the photons in modes $2$ and $3$, we evaluate the HOM visibility%
\begin{equation}
    V_{HOM}=\frac{\mathcal{P}(\infty)-\mathcal{P}(0)}{\mathcal{P}(\infty)},\label{homvis}
\end{equation}
where $\mathcal{P}(\infty)$ gives the probability of the coincidence measurement when the two wave packets are infinitely separated in time and do not interact. In practice, the time delay must be much larger than the photon's coherence time.

\begin{figure}
\includegraphics[width=.48\textwidth]{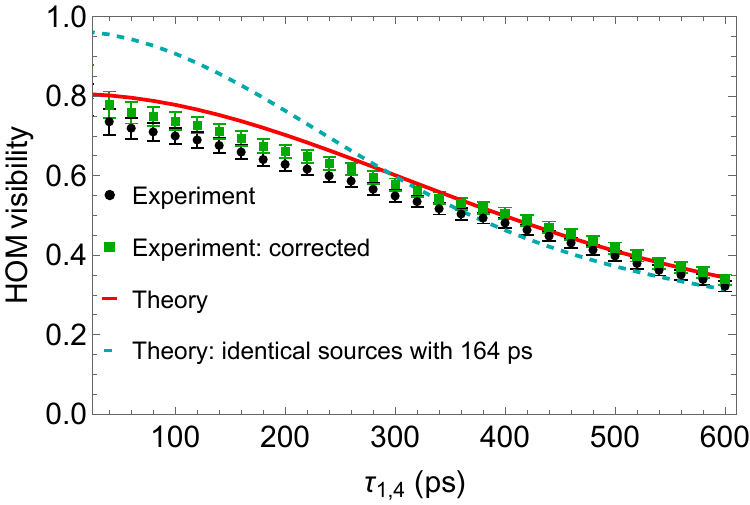}
\caption{HOM visibility as a function of the coincidence window $\tau_{1,4}$. The theoretical values have been calculated with Eqs. \eqref{HOMmain} and \eqref{homvis} by using the parameters presented in Fig. \ref{HOMcurves}. The black circles correspond to raw data, while the green squares represent the corrected HOM visibility after subtracting the accidental counts. The dashed blue line represents the theoretical calculation for the scenario where the two sources are identical with $T_{c,1}=T_{c,2}=165$\,ps. The theoretical prediction of HOM visibility in this scenario is $0.956$ for $\tau_{1,4}=40$\,ps.} \label{fig3}
\end{figure}

Figure \ref{fig3} shows the measured (black dots) and predicted (red curve) visibility as a function of $\tau_{1,4}$. There are two key observations. First, the visibility is below unity because the photons were partially distinguishable due to different coherence times of the two sources. A simulation for identical sources ($T_{c,1}=T_{c,2}=164$\,ps.) is shown for comparison (blue dashed curve). Second, the raw visibilities fall below the theoretical prediction because the theory assumes the emission of exactly two photon pairs, omitting higher-order pair emissions, stray photons, or dark counts. After subtracting accidental coincidence contributions from experimental data (green squares), the corrected visibilities follow the theoretical prediction (see Supplementary material for information on the quantification of accidental counts).
\subsection{Interplay among HOM visibility, coherence time, and timing jitter}
Having validated the model against the fourfold-HOM measurements, we now use it to explore the parameter trade-offs governing CW interference. For this analysis, we consider identical sources with rectangular spectral filters and neglect accidental contributions in order to isolate effects arising from temporal distinguishability. We compute the HOM visibility as a function of coherence time $T_c$ and coincidence window $\tau_{1,4}$ for fixed detector timing jitter.

\begin{figure}
\includegraphics[width=.47\textwidth]{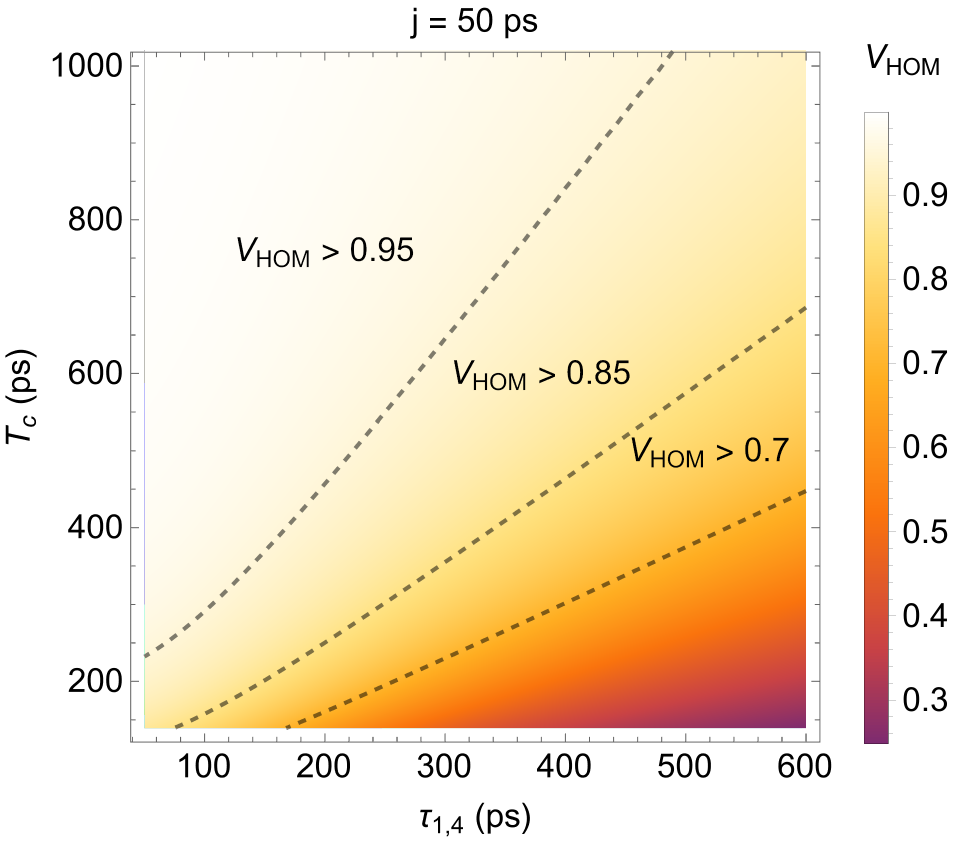}
\caption{Theoretical calculation of HOM visibility as a function of coherence time $T_c$ and coincidence window $\tau_{1,4}$ for fixed timing jitter $j=50$\,ps. The visibility is approximately constant for fixed ratios $T_c/\tau_{1,4}$. The deviation from linear behavior at short timescales arises from the finite timing jitter.} \label{fig5}
\end{figure}

Figure~\ref{fig5} shows that, for a given target visibility, the scaling is primarily determined by the ratio $T_c/\tau_{1,4}$, with small deviation at small timescales arising from finite timing jitter. For example, achieving $V_{HOM}=0.95$ requires $T_c/\tau_{1,4}\gtrsim 3.5$ for coherence times up to $250$\,ps, while visibilities around $0.85$ are obtained already for $T_c/\tau_{1,4}\approx 1$. In practice, this ratio must be chosen more conservatively to mitigate the impact of accidental coincidences. In Sec.~\ref{sec5}, we use this framework to optimize the four-photon rate with respect to $T_c$, $\tau_w$, and $j$.%

We reiterate that the method for estimating the HOM visibility shown in Fig.~\ref{window} (i.e, the condition $\tau_{2,3} \gg T_c$) evaluates the on-demand visibility of the heralded photons in modes $2$ and $3$. This estimation relies exclusively on the temporal filtering of the heralding photons in modes $1$ and $4$ (for more detailed explanation see the appendix \ref{appen1}). The HOM visibility can be further improved by applying temporal filtering to photons $2$ and $3$ as well (i.e., using narrow coincidence windows). Doing so allows the required ratio $T_c/\tau_{1,4}$ for a given target visibility to be relaxed by up to a factor of two. However, a different method must be applied to estimate the HOM visibility in this regime. A detailed discussion of that calibration procedure will be provided in a follow-up work.

In an entanglement swapping experiment, the roles of heralding and heralded photons are reversed: the coincidence detection of photons in modes $2$ and $3$ (the Bell state measurement) heralds the entangled state in modes $1$ and $4$. Consequently, the coincidence window $\tau_{2,3}$ must be narrow, while $\tau_{1,4}$ can remain wide.

\begin{figure}
\includegraphics[width=.5\textwidth]{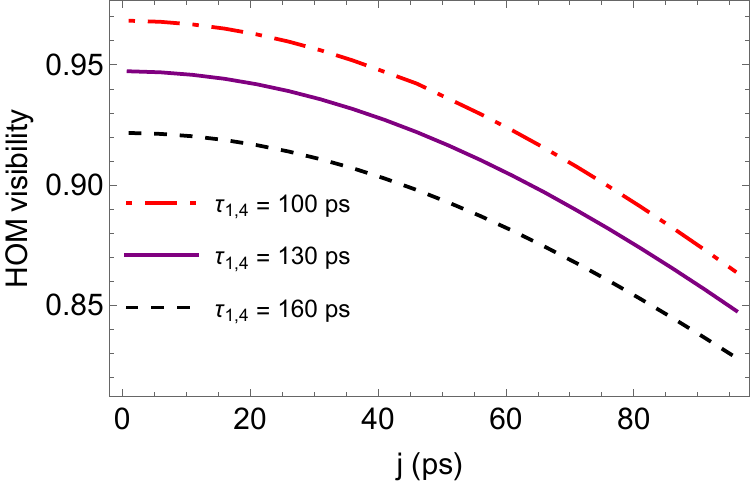}
\caption{Theoretical calculation of HOM visibility as a function of the timing jitter $j$ for different coincidence windows $\tau_{1,4}$. A rectangular filter function was applied, which results in a coherence time of $285$ ps for both sources. The visibility can be increased by decreasing the timing jitter. The accidental counts are not included in this calculation. } \label{fig6}
\end{figure}

Beyond optimization of the coincidence window, HOM visibility can be enhanced by decreasing timing jitter. Figure \ref{fig6} shows the theoretical visibility as a function of timing jitter $j$ for different coincidence windows with a coherence time of $285$ ps. While decreasing $j$ systematically improves the visibility, a finite upper bound remains for each coincidence window. This limit arises from residual temporal distinguishability introduced by the finite coincidence window.
\subsection{Rate-visibility trade-off}\label{sec5}

\subsubsection{Maximal four-photon rate}
The preceding analysis shows that achieving high HOM visibility requires reducing the coincidence window $\tau_w$. This improves temporal indistinguishability but simultaneously reduces the number of usable events.

A key limitation of SPDC sources is multi-pair emission, i.e., the generation of more than one photon pair per source within a single coherence time $T_c$ \cite{TAKESUE2010276}. Suppressing this noise requires operation in the photon-starved regime $\mu\ll1$, where $\mu$ denotes the mean number of generated pairs per pump pulse for pulsed operation, or per coherence time for CW operation. A typical choice is $\mu \approx 0.01$. In the CW regime, the effective temporal mode rate is set by $1/T_c$, and the average pair-generation rate of a single source is therefore $\mu/T_c$. Here, $T_c$ is the coherence time of identical sources.

Entanglement swapping with independent CW sources requires that both sources emit a pair within the same coincidence window $\tau_w$. The probability that a source emits a pair inside this window is $\mu\,\tau_w/T_c$. For two identical and independent sources, the probability that both emit a pair within the same window is therefore $(\mu\,\tau_w/T_c)^2$. Dividing by the window duration $\tau_w$ therefore yields the corresponding four-photon coincidence rate,
\begin{equation}
    R=   \left(\frac{1}{T_c} \mu\right)^2 \tau_w.\label{rate1}
\end{equation}
Equation~\eqref{rate1} describes the obtainable four-photon rate in the absence of transmission and detection loss.%
%
%
%
%
%
\begin{figure}
\includegraphics[width=.5\textwidth]{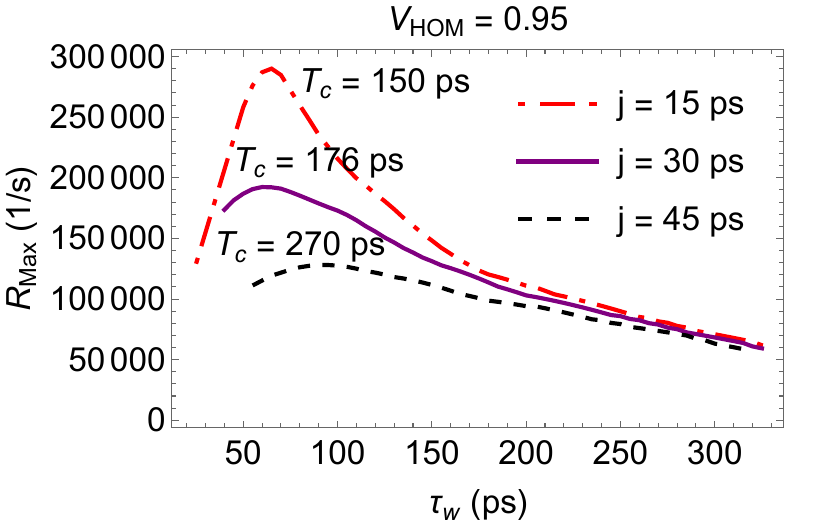}
\caption{Maximum achievable four-photon rate per second as a function of coincidence window for various timing jitters and for a fixed HOM visibility $V_{HOM}=0.95$. We observe that a reduction in timing jitter allows us to reach a higher photon rate. The coherence time is varied with $\tau_w$ to maintain the required HOM visibility; the depicted coherence times $T_c$ correspond to values at which the rates are maximized.}\label{fig11}
\end{figure}
\begin{figure}
\includegraphics[width=.5\textwidth]{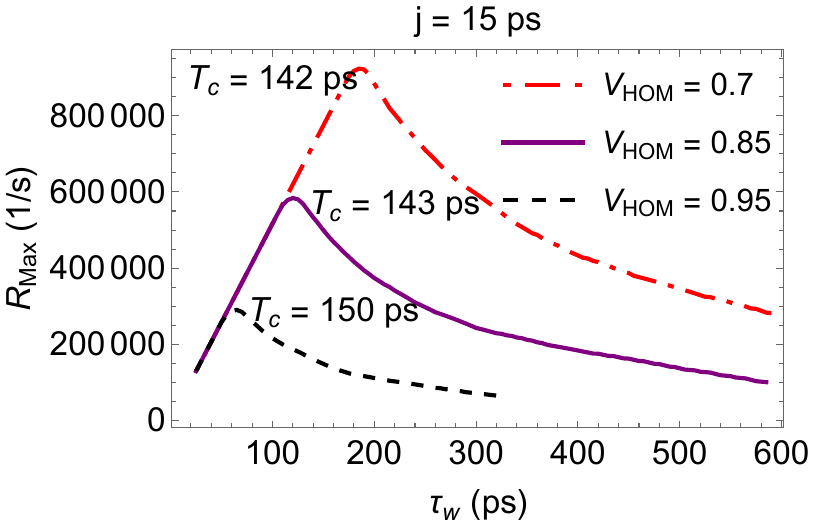}
\caption{Same as Fig. \ref{fig11}, but for various HOM visibilities and for fixed timing jitter $j=15$\, ps. There is an optimal size of $\tau_w$ for each $V_{HOM}$, which maximizes the four-photon rate. The depicted $T_c$ values correspond to values at which the rates are maximized.}\label{fig12}
\end{figure}
We now use this scaling to determine the coincidence window that maximizes the usable four-photon rate for a given target visibility. Figure~\ref{fig11} shows the maximal four-photon rates as a function of the coincidence window $\tau_w$ for different detector timing jitters. For each point, the coherence time is adjusted to maintain the target HOM visibility. We set the mean photon number to $\mu=0.01$, ensuring low multi-pair contributions. The coherence time is limited to $800$\,ps to remain within realistic source parameters. An optimal coincidence window clearly emerges for each jitter value, demonstrating that the usable rate can be maximized by jointly choosing $\tau_w$ and $T_c$ for a desired visibility. Figure~\ref{fig12} shows the same optimization for fixed timing jitter and varying target visibility, where we notice that a reduction of the target visibility can drastically increase the maximum detection rate.
\subsubsection{Comparison of pulsed and CW regimes}
It is instructive to compare the timing constraints governing indistinguishability in pulsed and CW SPDC sources. This comparison is intended as an order-of-magnitude scaling estimate under matched indistinguishability constraints and does not include implementation-specific prefactors.

In the pulsed regime, photon-pair creation is confined to a pump pulse of duration $\tau_p$, while consecutive pulses are separated by the inverse repetition rate $1/f_{\mathrm{rep}}$. To avoid cross-talk between neighboring pulses, the coincidence window must satisfy $\tau_w \lesssim 1/f_{\mathrm{rep}}$. Moreover, the single-photon coherence time $T_c$ should be comparable to or exceed the pump pulse duration $\tau_p$ to suppress timing distinguishability within a pulse \cite{PhysRevLett.94.083601}. Therefore, the degree of single temporal-mode selection is governed by the dimensionless ratio $\tau_p/T_c$. Under indistinguishability constraints comparable to those in the CW case, the per-pulse four-photon probability scales as $(\mu_p\,\tau_p/T_c)^2$, where $\mu_p$ is the per-pulse probability to generate a photon pair. Multiplication of the four-photon probability by the repetition rate yields the four-photon coincidence rate
\begin{equation*}
    R_{\mathrm{pulsed}} \approx (\mu_p\,\tau_p/T_c)^2 f_{\mathrm{rep}},
\end{equation*}
up to constant factors such as Bell-state measurement efficiency and transmission or detection losses. 

Here, two primary comparisons are necessary. First, as shown in Fig. \ref{fig3}, achieving HOM visibilities exceeding 90\% in the CW regime requires the ratio $T_c/\tau_w$ to be approximately greater than $1.5$. A comparable ratio of the coherence time to the pulse duration, $T_c/\tau_p$, is experimentally accessible in the pulsed regime.

Second, we must compare the effective temporal mode rates by evaluating the equivalence $f_{\mathrm{rep}} \sim 1/\tau_w$. GHz-class pulsed laser and SPDC source platforms relevant for high-rate photonic interference protocols have been demonstrated \cite{Keller2003,Wakui:20,Tsujimoto:21,Cabrejo-Ponce_2022}. For repetition rates between $1$ GHz to $50$ GHz system, the temporal separation between pulses ranges from 1 ns down to 20 ps. Coincidence windows of this duration are within the reach of state-of-the-art timing systems, a capability we also leveraged in our experiment. Therefore, by employing comparable temporal selection criteria, the achievable four-photon rates in the pulsed and CW regimes become comparable.

If we assume a $1$ GHz repetition-rate, $\tau_p \sim T_c$ and $\mu_p\approx0.01$, the maximal four-fold rate is bounded by $\mu_p^2 \,f_{\mathrm{rep}}$ that corresponds to the order of $10^5\,\mathrm{s}^{-1}$. This is comparable to the rates predicted in Figs.~\ref{fig11} and~\ref{fig12} for the CW regime. Note that if higher repetition rates are used, the pulsed regime may exceed the CW regime in terms of rate.

The key distinction, however, is that CW operation removes the need for optical emission-time and path-length stabilization between independent sources. Instead, only electronic time-stamping and classical clock synchronization are required, shifting the synchronization burden from the optical to the electronic domain, which is considerably more stable and scalable.

Overall, CW operation trades a moderate reduction in maximum achievable rate for a substantial relaxation of optical synchronization requirements. This trade-off might become favorable for long-distance quantum networking, where stabilizing optical paths between independent photon sources is technically demanding, in particular in the context of satellite links. In practice, the spectral filtering required to reach long coherence times can introduce significant additional optical loss, which further motivates the development of intrinsically narrowband or spectrally engineered photon-pair sources.

\subsection{Application to long-distance quantum-network links}%
We now apply the rate-visibility analysis of asynchronous entanglement swapping to a scenario operating under realistic transmission loss. Such loss-limited regimes are characteristic of long-distance quantum networks, including satellite links, where optical synchronization between independent photon sources becomes increasingly challenging. In the pulsed regime, this synchronization must be maintained at the level of the optical pump pulses, i.e., typically on sub-picosecond timescales. The asynchronous CW approach is therefore particularly well suited to this regime.
\begin{figure}
\includegraphics[width=.45\textwidth]{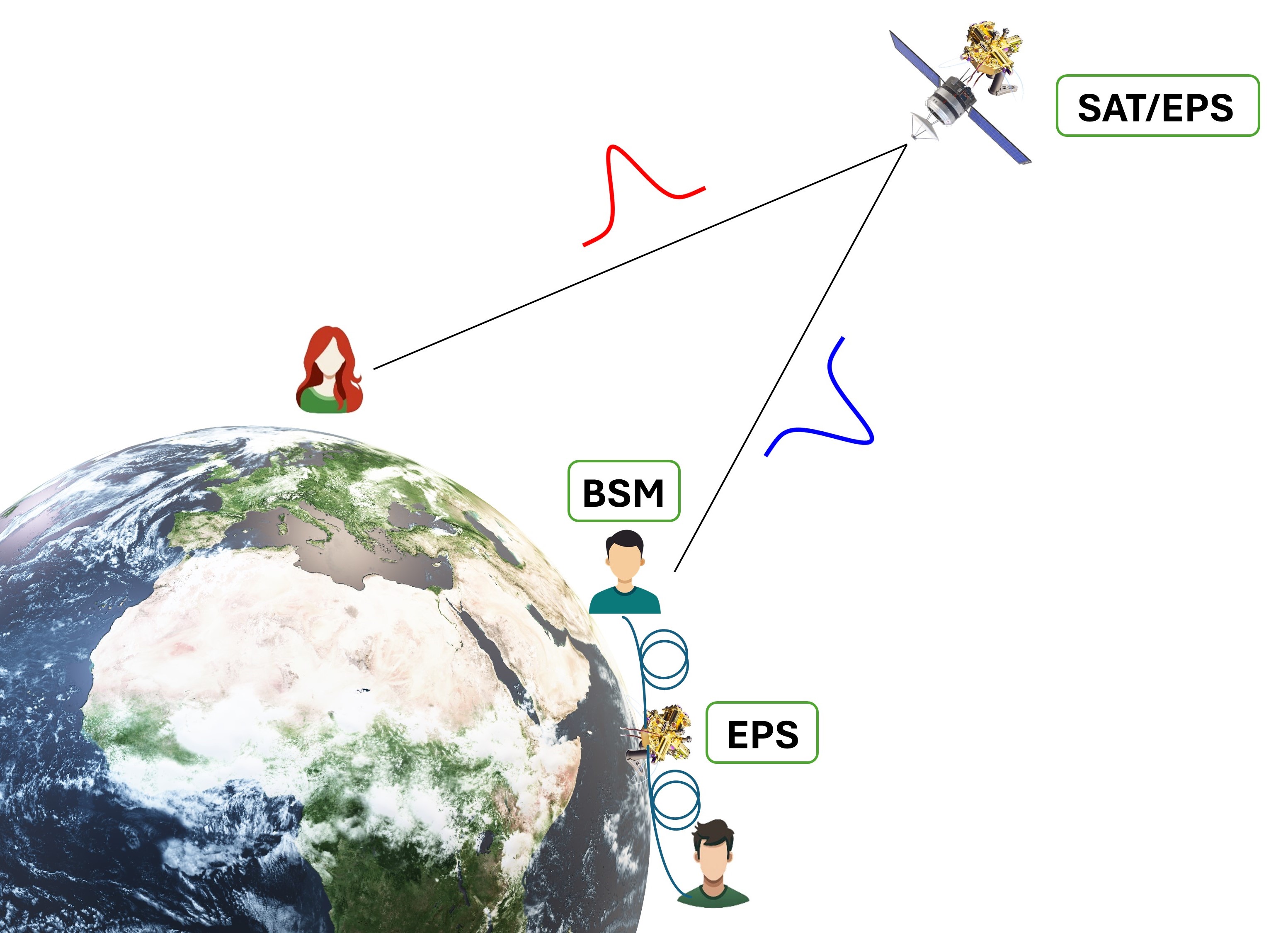}
\caption{Example high-loss network scenario: entanglement swapping with a Bell-state measurement (BSM) performed at a ground station. Entangled photon pairs are generated by a space-based entangled photon source (EPS) onboard a satellite and by a ground-based EPS, while the BSM links free-space and fiber segments.}\label{fig13}
\end{figure}

Including transmission and detection efficiencies $\eta_i$, the estimated four-photon coincidence rate becomes%
\begin{equation}
   R= \left(\frac{1}{T_c} \mu\right)^2 \tau_w \times \frac{1}{2} \times \eta_1\eta_2\eta_3\eta_4,\label{rate2}
\end{equation}
where the factor $1/2$ accounts for the intrinsic success probability of linear-optical Bell-state measurements.

As a concrete example of a high-loss scenario, we consider a dual-downlink satellite geometry in which an entangled photon source is located onboard a satellite and the Bell-state measurement is performed at a ground station as shown in Fig. \ref{fig13}. Assuming realistic values for sender and receiver telescopes, adaptive optics, atmospheric turbulence, and pointing and tracking, we simulate downlink losses via a numerical framework described in \cite{AndrejCode} at $\lambda=1550\,$nm. We simulate the total amount of passes with simultaneous access over the course of one year for ground stations in London (Alice) and Berlin (Bob, BSM), a satellite in the orbit of the Micius mission \cite{Bedington2017}, and link distances of 10\,km (with losses of 2.2\,dB) from the on-ground EPS to Bob and Charlie respectively. Especially, we integrate Eq.~\eqref{rate2} to find the amount of swaps per pass. We assume $\mu=0.01$, filter bandwidths of 10\,pm and a coincidence window of 50\,ps, which should ensure a high HOM visibility. The losses for the median pass range between roughly 30 and 24\,dB during the acquisition window. For these values, we find 872 swapping events over a year spread out between 316 passes, with a median swapping rate of $\approx 3$ swapping events per pass. Notably, this scenario does not require optical memories or active synchronization of photon emission times between the two sources, both of which represent major technical challenges for long-distance quantum networking.

The resulting swapping rates remain modest in this high-loss regime but can be increased by incorporating multiplexing strategies and operating at higher brightness, deterministic pair sources, as well as further developments in quantum photonics for higher repetition rates and faster detectors. 

These results highlight that the main advantage of CW operation in such scenarios lies not in raw rate improvements, but in the removal of stringent optical synchronization requirements, which become increasingly difficult in long-distance quantum networks involving satellites.

\section{conclusion}
We developed and experimentally validated a theoretical framework that quantitatively describes multi-photon interference in the CW SPDC regime. The model incorporates detector timing jitter $j$, photon coherence time $T_c$, and the coincidence window $\tau_w$ into an effective description of temporal indistinguishability, and provides a consistent procedure to estimate HOM visibility in asynchronous applications, particularly relevant to long-distance quantum communication networks.

Using this framework, we identified the scaling laws governing time-resolved HOM interference. We showed that the visibility is primarily determined by the ratio $T_c/\tau_w$, corresponding to an effective temporal-mode number, with finite detector timing jitter introducing a small deviation for short timescales. 

Finally, we translated these insights into practical design rules for asynchronous quantum networking. For a given target visibility and fixed timing jitter, an optimal coincidence window exists that maximizes the usable four-photon rate. Comparison with pulsed SPDC sources shows that CW operation can achieve comparable rates while removing the need for optical emission-time synchronization between independent sources. This relaxation of synchronization requirements makes the CW approach particularly attractive for scalable long-distance quantum networks.

\section*{Acknowledgements}
We thank Carlos Sevilla-Gutiérrez for helpful discussions, Christopher Spiess for the code to perform post-selection of time tags.

\section*{Funding}
 ML, SH, SN, and FS are part of the Max Planck School of Photonics supported by BMFTR, Max Planck Society, and Fraunhofer Society. This work was co-funded by the Natural Sciences and Engineering Research Council of Canada (NSERC) and the European Commission under Grant no. 101070168 (HyperSpace) and the Carl-Zeiss-Stiftung within the Carz-Zeiss-Stiftung Center for Quantum Photonics (CZS QPhoton) under the project ID P2021-00-019.

\section*{Data Availability Statement}
The data that support the findings of this study are available from the corresponding authors upon reasonable request.

\appendix
\section{The role of the coincidence window of BS photons}\label{appen1}
\begin{figure}
\includegraphics[width=.48\textwidth]{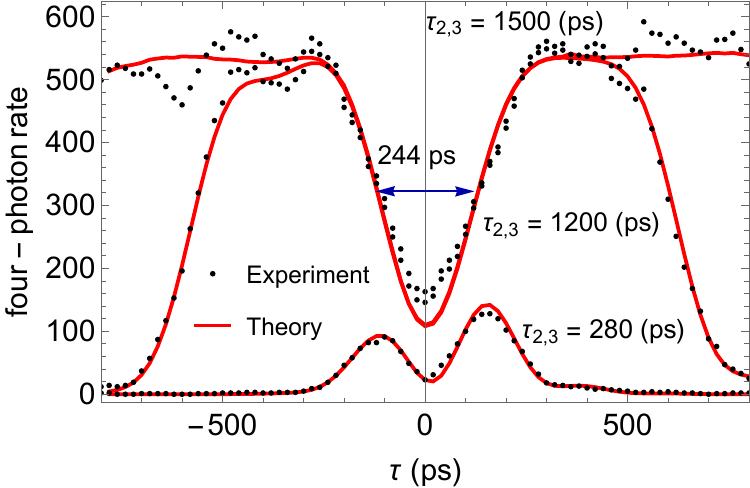}
\caption{HOM measurement for different coincidence windows between BS photons $\tau_{2,3}$ for fixed $\tau_{1,4}=80$\,ps. The width of the plateau region depends on $\tau_{2,3}$. The typical HOM curve is recovered by applying a coincidence window $\tau_{2,3}$, which is much larger than $T_c$.}\label{fig4}
\end{figure}

Figure \ref{fig4} illustrates four-photon counts for different $\tau_{2,3}$, but for fixed $\tau_{1,4}=80$\,ps, where the dots correspond to the experimental data and the solid curves are the corresponding theoretical predictions. The curve corresponding to $\tau_{2,3}=280$\,ps exhibits the characteristic local minimum at $\tau=0$. As the detection delay increases, the coincidence rate initially rises, reaches a maximum, and subsequently decreases to zero for time delays that exceed $T_c+\tau_{2,3}$.

This decrease is a direct consequence of the small coincidence windows in the spatial modes $2^{\prime}$ and $3^{\prime}$, which act as temporal filters. Due to the inherent time correlation between photons generated in pairs (e.g., $1$ with $2$, and $3$ with $4$), a large detection delay $\tau$ for photon $4$ implies a similar temporal shift for its partner photon. When $\tau$ exceeds the coherence time, this shift causes one of the interfered photons (either in the spatial modes $2^{\prime}$ or $3^{\prime}$) to fall outside the narrow coincidence window, thus preventing the detection of a four-photon coincidence event. This process resembles a time scan across the coherence time of interfering photons.

When $\tau_{2,3}$ is much larger than the coherence time $T_c$, the width of the plateau region is dictated by the size of $\tau_{2,3}$. The width of the HOM dip corresponding to $\tau_{2,3}=2000$\, ps is determined by the largest coherence time among the sources.

With the method used here, the condition $\tau_{2,3} \gg T_c$ must hold to obtain an unbiased estimate of the HOM visibility. It provides the indistinguishability photons without additional temporal filtering in spatial modes $2^{\prime}$ and $3^{\prime}$. However, when the noise level is high, using large coincidence windows $\tau_{2,3}$ can lead to a lower HOM visibility.

To avoid a high noise level, small coincidence windows can be applied in the BS channels as well, as shown in the $\tau_{2,3}=280$\,ps curve in Fig. \ref{fig4}. This acts as a temporal post-filter on already interfered photons, which will increase the measured visibility. In that case, our standard visibility evaluation no longer applies: while $\mathcal{P}(0)$ is measured correctly, the plateau value no longer represents the true distinguishable rate $\mathcal{P}(\infty)$.

To find the correct reference in this small-window regime, $\mathcal{P}(\infty)$ must be re-evaluated. One way is to apply a large delay $\tau$ to both channels $1$ and $2^{\prime}$ (or $1$ and $3^{\prime}$) simultaneously. The resulting rate at this large $\tau$ corresponds to the correct distinguishable rate for non-interfering pairs. A detailed discussion of this calibration procedure will be provided in a follow-up work. 
\section{Noise contribution of accidental coincidence counts}\label{accSec}
\begin{figure}
\includegraphics[width=.5\textwidth]{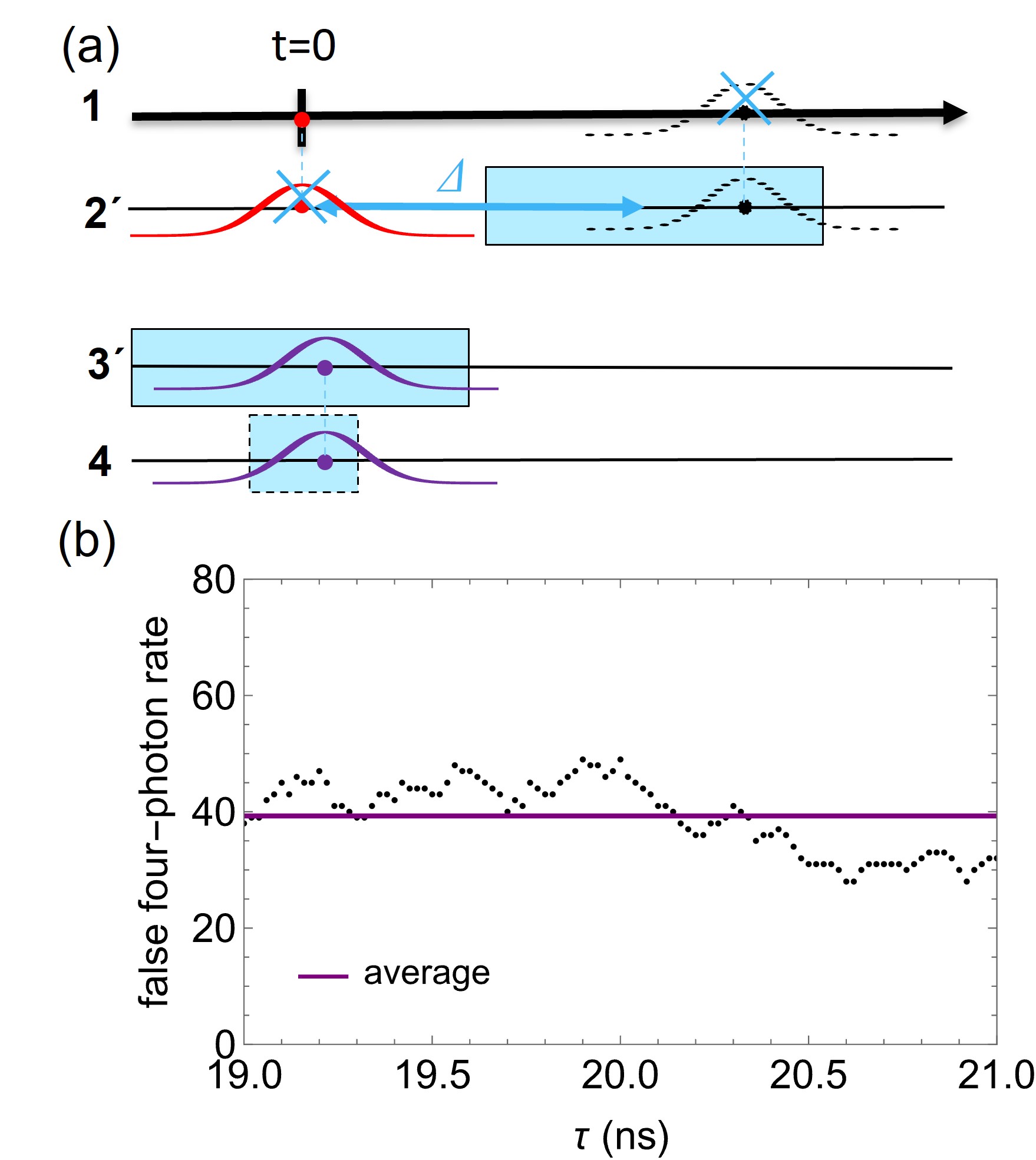}
\caption{Accidental coincidence counts are estimated by deliberately breaking the genuine pairwise time correlations. (a) The four-photon events contain at least one noise photon if a constant offset $\Delta$ is added to the time tags of spatial mode $2^{\prime}$. Instead of the genuine pair, the stray photon, presented with a dashed curve, is detected in the spatial mode $2^{\prime}$. The pairs of the trigger and the noise photons are not detected. (b)}\label{acc}
\end{figure}
To estimate the accidental four-photon coincidence probability and corresponding rate, we analyze all possible four-photon events and their detection probabilities. We first consider two independent sources in the absence of interference at the BS. Let $\mu_{c,1}$ and $\mu_{c,2}$ be the probabilities that first and second sources each emit a photon pair within the coincidence window $\tau_w$, and let $\eta_n$ be the detection probability for a photon in channel $n$. The probability that a pair from the first source is detected is given by $\mu_{c,1} \eta_1 \eta_2$ and for the second source by $\mu_{c,2} \eta_3 \eta_4$. Because the sources are independent, the probability to detect one pair from each source (a four-fold event) is the product of the individual probabilities: $\mu_{c,1} \eta_1 \eta_2 \cdot \mu_{c,2} \eta_3 \eta_4$.

When the interference at the BS is introduced, the four-fold probability is reduced compared to $\mu_{c,1} \eta_1 \eta_2 \cdot \mu_{c,2} \eta_3 \eta_4$. This is because the BS photons do not always exit into distinct output modes. We define the probability that the photons exit into the same spatial mode as $\gamma$. The probability that the photons exit into distinct spatial modes is $(1-\gamma)$. For compactness, $\eta_2$ and $\eta_3$ below refer to the detection efficiencies of channels $2^{\prime}$ and $3^{\prime}$. Therefore, the probability of four-photon coincidence is reduced by a factor of $(1-\gamma)$ and reads as $(1-\gamma) \mu_{c,2}\,\eta_4 \,\eta_3\,\mu_{c,1} \eta_1 \,\eta_2$. We refer to these four-photon events as real four-fold events, since they arise from exactly two pairs.

Apart from real four-fold events, accidental four-photon coincidences can arise from higher-order SPDC emission, i.e., when one source emits more than one pair within the same coincidence window. For instance, if the BS photons exit into the same spatial mode, one BS channel remains empty and can be occupied with a noise photon, as shown in Fig. \ref{window} (b). The probabilities that the two photons exit together into channels $2^{\prime}$ or $3^{\prime}$ are $\frac{\gamma}{2} \mu_{c,2}\,\eta_4\,\mu_{c,1} \eta_1\, \bar{\eta}_{2}$ and $\frac{\gamma}{2} \mu_{c,2}\,\eta_4\,\mu_{c,1} \eta_1\, \bar{\eta}_{3}$, respectively. The efficiency $\bar{\eta}_n =1-(1-\eta_n)^2$ denotes the probability that at least one of the two photons entering the channel $n$ is detected. Let $P_n$ denote the probability of an uncorrelated (noise) single count in channel $n$ within the coincidence window. The probabilities for false four-photon detection events become: $\frac{\gamma}{2} \mu_{c,2}\,\eta_4\,\mu_{c,1} \eta_1\, \bar{\eta}_{2}\, P_{3^{\prime}}$ and $\frac{\gamma}{2} \mu_{c,2}\,\eta_4\,\mu_{c,1} \eta_1\, \bar{\eta}_{3}\, P_{2^{\prime}}$. 

Another category of false four-photon events consists of one real pair plus two noise photons. For example, the detection of a pair from source 2 (channels $3^{\prime}$ and 4) together with noise photons in channels 1 and $2^{\prime}$ occurs with probability $0.5\,\mu_{c,2}\,\eta_4 \,\eta_3\,P_1\,\,P_{2^{\prime}}$. If we account for all four “pair + two noise photons” combinations, the total four-photon probability at $\tau=0$ becomes
\begin{align*}
A_0 &= (1-\gamma) \mu_{c,2}\,\eta_4 \,\eta_3\,\mu_{c,1} \eta_1 \,\eta_2  + \frac{\gamma}{2} \mu_{c,2}\,\eta_4\,\mu_{c,1} \eta_1\, \bar{\eta}_{2}\, P_{3^{\prime}}\\
& + \frac{\gamma}{2} \mu_{c,2}\,\eta_4\,\mu_{c,1} \eta_1\, \bar{\eta}_{3}\, P_{2^{\prime}} + \frac{1}{2}\mu_{c,2}\,\eta_4 \,\eta_3\,P_1\,\,P_{2^{\prime}} \\&+
\frac{1}{2}\mu_{c,2}\,\eta_4 \,\eta_2\,P_1\,\,P_{3^{\prime}}+ P_4\,P_{3^{\prime}}\frac{1}{2}\mu_{c,1}\,\eta_1 \,\eta_2\\&+ P_4\,P_{2^{\prime}}\frac{1}{2}\mu_{c,1}\,\eta_1 \,\eta_3,
\end{align*}
where only the first term represents the real four-photon events. To validate our noiseless model, we can subtract the remaining terms from the raw counts and check whether the result agrees with the corrected data. In an actual entanglement-swapping experiment, however, this subtraction is generally not available in real time. Events with four uncorrelated photons are, in principle, possible but have negligibly small probability. 

To estimate the four-photon accidental background, we deliberately break the genuine pairwise correlations so that the surviving four-photon events include at least one noise photon. We do this by adding a constant offset $\Delta \gg \tau_{2,3}$, first to all time tags of spatial mode $2^{\prime}$ as illustrated in Fig. \ref{acc} (a). This decorrelates the spatial mode $2^{\prime}$ from the other detections. Under this condition, any observed four-photon event must include a noise photon in the channel $2^{\prime}$ with probability $P_{2^{\prime}}$. This includes "pair + two noise photon" events (e.g., one real pair from the second source and two noise photons, one of which is $P_{2^{\prime}}$). Moreover, the probability $\frac{\gamma}{2} \mu_{c,2}\,\eta_4\,\mu_{c,1} \eta_1\, \bar{\eta}_{3}\, P_{2^{\prime}}$ is independent of the position of the channel $2^{\prime}$ and will therefore occur in the channel $2^{\prime}$-shifted rate. The probability of a four-photon event in this channel $2^{\prime}$-shifted scenario is given by 
\begin{align*}
    A_{s,2^{\prime}} &=\frac{1}{2}\mu_{c,2}\,\eta_4 \,\eta_3\,P_1\,\,P_{2^{\prime}}+ P_4\,P_{2^{\prime}}\frac{1}{2}\mu_{c,1}\,\eta_1 \,\eta_3\\& +\frac{\gamma}{2} \mu_{c,2}\,\eta_4\,\mu_{c,1} \eta_1\, \bar{\eta}_{3}\, P_{2^{\prime}}.
\end{align*}
Similarly, if the channel $3^{\prime}$ is shifted, we have
\begin{align*}
    A_{s,3^{\prime}} &=P_4\,P_{3^{\prime}}\frac{1}{2}\mu_{c,1}\,\eta_1 \,\eta_2+\frac{1}{2}\mu_{c,2}\,\eta_4 \,\eta_2\,P_1\,\,P_{3^{\prime}}\\&+\frac{\gamma}{2} \mu_{c,2}\,\eta_4\,\mu_{c,1} \eta_1\, \bar{\eta}_{2}\, P_{3^{\prime}}.
\end{align*}
Subtracting the single-shifted contributions $A_{s,2^{\prime}}$ and $A_{s,3^{\prime}}$ from the total four-photon probability $A_0$ yields:
\begin{equation*}
\mathcal{P}(0)=A_0-A_{s,2^{\prime}}-A_{s,3^{\prime}},
\end{equation*}
where $\mathcal{P}(0)$ denotes the probability of a real four-photon event, as defined in the main text. The experimental correction follows an analogous procedure: the raw counts at $\tau=0$ are first estimated, followed by the subtraction of the counts from the shifted time tags corresponding to $A_{s,2^{\prime}}$ and $A_{s,3^{\prime}}$. The estimation of $\mathcal{P}(\infty)$ is conducted in a similar manner. The advantage of this method is that no additional measurements are required, and the shifted rates corresponding to probabilities $A_{s,2^{\prime}}$ and $A_{s,3^{\prime}}$ can be estimated from the same data set. Figure \ref{acc} (b) shows the counts corresponding to $A_{s,2^{\prime}}$ with the average noise level for $\tau_{1,4}=600$\,ps and $\tau_{2,3}=2000$\,ps.

\bibliographystyle{apsrev4-1}
\bibliography{bibliography.bib}

@article{Cabrejo-Ponce_2022,
doi = {10.1088/2058-9565/ac86f0},
url = {https://doi.org/10.1088/2058-9565/ac86f0},
year = {2022},
month = {aug},
publisher = {IOP Publishing},
volume = {7},
number = {4},
pages = {045022},
author = {Cabrejo-Ponce, Meritxell and Spiess, Christopher and Marques Muniz, André Luiz and Ancsin, Philippe and Steinlechner, Fabian},
title = {GHz-pulsed source of entangled photons for reconfigurable quantum networks},
journal = {Quantum Science and Technology}
}

@article{PhysRevLett.128.063602,
  title = {Heralding Multiple Photonic Pulsed Bell Pairs via Frequency-Resolved Entanglement Swapping},
  author = {Merkouche, Sofiane and Thiel, Val\'erian and Davis, Alex O. C. and Smith, Brian J.},
  journal = {Phys. Rev. Lett.},
  volume = {128},
  issue = {6},
  pages = {063602},
  numpages = {5},
  year = {2022},
  month = {Feb},
  publisher = {American Physical Society},
  doi = {10.1103/PhysRevLett.128.063602},
  url = {https://link.aps.org/doi/10.1103/PhysRevLett.128.063602}
}

@article{PhysRevLett.96.110501,
  title = {Experimental Synchronization of Independent Entangled Photon Sources},
  author = {Yang, Tao and Zhang, Qiang and Chen, Teng-Yun and Lu, Shan and Yin, Juan and Pan, Jian-Wei and Wei, Zhi-Yi and Tian, Jing-Rong and Zhang, Jie},
  journal = {Phys. Rev. Lett.},
  volume = {96},
  issue = {11},
  pages = {110501},
  numpages = {4},
  year = {2006},
  month = {Mar},
  publisher = {American Physical Society},
  doi = {10.1103/PhysRevLett.96.110501},
  url = {https://link.aps.org/doi/10.1103/PhysRevLett.96.110501}   
}

@article{PhysRevLett.81.5932,
  title = {Quantum Repeaters: The Role of Imperfect Local Operations in Quantum Communication},
  author = {Briegel, H.-J. and D\"ur, W. and Cirac, J. I. and Zoller, P.},
  journal = {Phys. Rev. Lett.},
  volume = {81},
  issue = {26},
  pages = {5932--5935},
  numpages = {0},
  year = {1998},
  month = {Dec},
  publisher = {American Physical Society},
  doi = {10.1103/PhysRevLett.81.5932},
  url = {https://link.aps.org/doi/10.1103/PhysRevLett.81.5932}   
}

@article{PhysRevLett.90.207901,
  title = {Experimental Realization of Entanglement Concentration and a Quantum Repeater},
  author = {Zhao, Zhi and Yang, Tao and Chen, Yu-Ao and Zhang, An-Ning and Pan, Jian-Wei},
  journal = {Phys. Rev. Lett.},
  volume = {90},
  issue = {20},
  pages = {207901},
  numpages = {4},
  year = {2003},
  month = {May},
  publisher = {American Physical Society},
  doi = {10.1103/PhysRevLett.90.207901},
  url = {https://link.aps.org/doi/10.1103/PhysRevLett.90.207901}   
}

@Article{Duan2001,
author={Duan, L.-M.
and Lukin, M. D.
and Cirac, J. I.
and Zoller, P.},
title={Long-distance quantum communication with atomic ensembles and linear optics},
journal={Nature},
year={2001},
month={Nov},
day={01},
volume={414},
number={6862},
pages={413-418},
issn={1476-4687},
doi={10.1038/35106500},
url={https://doi.org/10.1038/35106500}
}

@article{doi:10.1139/cjp-2023-0312,
author = {Drago, Christian and Brańczyk, Agata M.},
title = {Hong–Ou–Mandel interference: a spectral–temporal analysis},
journal = {Canadian Journal of Physics},
volume = {102},
number = {8},
pages = {411-421},
year = {2024},
doi = {10.1139/cjp-2023-0312},

URL = { 
    
        https://doi.org/10.1139/cjp-2023-0312
    
    

},
eprint = { 
    
        https://doi.org/10.1139/cjp-2023-0312
    
    

}
}

@article{PhysRevLett.123.160502,
  title = {Entanglement Swapping with Semiconductor-Generated Photons Violates Bell's Inequality},
  author = {Zopf, Michael and Keil, Robert and Chen, Yan and Yang, Jingzhong and Chen, Disheng and Ding, Fei and Schmidt, Oliver G.},
  journal = {Phys. Rev. Lett.},
  volume = {123},
  issue = {16},
  pages = {160502},
  numpages = {7},
  year = {2019},
  month = {Oct},
  publisher = {American Physical Society},
  doi = {10.1103/PhysRevLett.123.160502},
  url = {https://link.aps.org/doi/10.1103/PhysRevLett.123.160502}
}

@article{PhysRevA.64.063815,
  title = {Eliminating frequency and space-time correlations in multiphoton states},
  author = {Grice, W. P. and U'Ren, A. B. and Walmsley, I. A.},
  journal = {Phys. Rev. A},
  volume = {64},
  issue = {6},
  pages = {063815},
  numpages = {7},
  year = {2001},
  month = {Nov},
  publisher = {American Physical Society},
  doi = {10.1103/PhysRevA.64.063815},
  url = {https://link.aps.org/doi/10.1103/PhysRevA.64.063815}
}

@article{Tambasco:16,
author = {J-L. Tambasco and A. Boes and L. G. Helt and M. J. Steel and A. Mitchell},
journal = {Opt. Express},
number = {17},
pages = {19616--19626},
publisher = {Optica Publishing Group},
title = {Domain engineering algorithm for practical and effective photon sources},
volume = {24},
month = {Aug},
year = {2016},
url = {https://opg.optica.org/oe/abstract.cfm?URI=oe-24-17-19616},
doi = {10.1364/OE.24.019616},
}

@article{PhysRevA.108.023718,
  title = {Enhancing the purity of single photons in parametric down-conversion through simultaneous pump-beam and crystal-domain engineering},
  author = {Baghdasaryan, Baghdasar and Steinlechner, Fabian and Fritzsche, Stephan},
  journal = {Phys. Rev. A},
  volume = {108},
  issue = {2},
  pages = {023718},
  numpages = {8},
  year = {2023},
  month = {Aug},
  publisher = {American Physical Society},
  doi = {10.1103/PhysRevA.108.023718},
  url = {https://link.aps.org/doi/10.1103/PhysRevA.108.023718}
}

@article{Pickston:21,
author = {Alexander Pickston and Francesco Graffitti and Peter Barrow and Christopher L. Morrison and Joseph Ho and Agata M. Bra\'{n}czyk and Alessandro Fedrizzi},
journal = {Opt. Express},
number = {5},
pages = {6991--7002},
publisher = {Optica Publishing Group},
title = {Optimised domain-engineered crystals for pure telecom photon sources},
volume = {29},
month = {Mar},
year = {2021},
url = {https://opg.optica.org/oe/abstract.cfm?URI=oe-29-5-6991},
doi = {10.1364/OE.416843},
}

@article{Graffitti:18,
author = {Francesco Graffitti and Peter Barrow and Massimiliano Proietti and Dmytro Kundys and Alessandro Fedrizzi},
journal = {Optica},
number = {5},
pages = {514--517},
publisher = {Optica Publishing Group},
title = {Independent high-purity photons created in domain-engineered crystals},
volume = {5},
month = {May},
year = {2018},
url = {https://opg.optica.org/optica/abstract.cfm?URI=optica-5-5-514},
doi = {10.1364/OPTICA.5.000514},
}

@article{PhysRevA.93.013801,
  title = {Shaping the joint spectrum of down-converted photons through optimized custom poling},
  author = {Dosseva, Annamaria and Cincio, \L{}ukasz and Bra\ifmmode \acute{n}\else \'{n}\fi{}czyk, Agata M.},
  journal = {Phys. Rev. A},
  volume = {93},
  issue = {1},
  pages = {013801},
  numpages = {7},
  year = {2016},
  month = {Jan},
  publisher = {American Physical Society},
  doi = {10.1103/PhysRevA.93.013801},
  url = {https://link.aps.org/doi/10.1103/PhysRevA.93.013801}
}

@article{BenDixon:13,
author = {P. Ben Dixon and Jeffrey H. Shapiro and Franco N. C. Wong},
journal = {Opt. Express},
number = {5},
pages = {5879--5890},
publisher = {Optica Publishing Group},
title = {Spectral engineering by Gaussian phase-matching for quantum photonics},
volume = {21},
month = {Mar},
year = {2013},
url = {https://opg.optica.org/oe/abstract.cfm?URI=oe-21-5-5879},
doi = {10.1364/OE.21.005879},
}

@article{Branczyk:11,
author = {Agata M. Bra\'{n}czyk and Alessandro Fedrizzi and Thomas M. Stace and Tim C. Ralph and Andrew G. White},
journal = {Opt. Express},
number = {1},
pages = {55--65},
publisher = {Optica Publishing Group},
title = {Engineered optical nonlinearity for quantum light sources},
volume = {19},
month = {Jan},
year = {2011},
url = {https://opg.optica.org/oe/abstract.cfm?URI=oe-19-1-55},
doi = {10.1364/OE.19.000055},
}

@article{PhysRevA.96.053842,
  title = {Effects of filtering on the purity of heralded single photons from parametric sources},
  author = {Blay, Daniel R. and Steel, M. J. and Helt, L. G.},
  journal = {Phys. Rev. A},
  volume = {96},
  issue = {5},
  pages = {053842},
  numpages = {8},
  year = {2017},
  month = {Nov},
  publisher = {American Physical Society},
  doi = {10.1103/PhysRevA.96.053842},
  url = {https://link.aps.org/doi/10.1103/PhysRevA.96.053842}
}

@article{PhysRevA.101.063821,
  title = {Optimizing spontaneous parametric down-conversion sources for boson sampling},
  author = {van der Meer, R. and Renema, J. J. and Brecht, B. and Silberhorn, C. and Pinkse, P. W. H.},
  journal = {Phys. Rev. A},
  volume = {101},
  issue = {6},
  pages = {063821},
  numpages = {9},
  year = {2020},
  month = {Jun},
  publisher = {American Physical Society},
  doi = {10.1103/PhysRevA.101.063821},
  url = {https://link.aps.org/doi/10.1103/PhysRevA.101.063821}
}

@article{PhysRevLett.100.133601,
  title = {Heralded Generation of Ultrafast Single Photons in Pure Quantum States},
  author = {Mosley, Peter J. and Lundeen, Jeff S. and Smith, Brian J. and Wasylczyk, Piotr and U'Ren, Alfred B. and Silberhorn, Christine and Walmsley, Ian A.},
  journal = {Phys. Rev. Lett.},
  volume = {100},
  issue = {13},
  pages = {133601},
  numpages = {4},
  year = {2008},
  month = {Apr},
  publisher = {American Physical Society},
  doi = {10.1103/PhysRevLett.100.133601},
  url = {https://link.aps.org/doi/10.1103/PhysRevLett.100.133601}
}

@article{Garay-Palmett:07,
author = {K. Garay-Palmett and H. J. McGuinness and Offir Cohen and J. S. Lundeen and R. Rangel-Rojo and A. B. U'Ren and M. G. Raymer and C. J. McKinstrie and S. Radic and I. A. Walmsley},
journal = {Opt. Express},
number = {22},
pages = {14870--14886},
publisher = {Optica Publishing Group},
title = {Photon pair-state preparation with tailored spectral properties by spontaneous four-wave mixing in photonic-crystal fiber},
volume = {15},
month = {Oct},
year = {2007},
url = {https://opg.optica.org/oe/abstract.cfm?URI=oe-15-22-14870},
doi = {10.1364/OE.15.014870},
}

@Article{Wang2021,
author={Wang, Weilong
and Tamaki, Kiyoshi
and Curty, Marcos},
title={Measurement-device-independent quantum key distribution with leaky sources},
journal={Scientific Reports},
year={2021},
month={Jan},
day={18},
volume={11},
number={1},
pages={1678},
issn={2045-2322},
doi={10.1038/s41598-021-81003-2},
url={https://doi.org/10.1038/s41598-021-81003-2}
}

@article{PhysRevLett.101.080403,
  title = {Multistage Entanglement Swapping},
  author = {Goebel, Alexander M. and Wagenknecht, Claudia and Zhang, Qiang and Chen, Yu-Ao and Chen, Kai and Schmiedmayer, J\"org and Pan, Jian-Wei},
  journal = {Phys. Rev. Lett.},
  volume = {101},
  issue = {8},
  pages = {080403},
  numpages = {4},
  year = {2008},
  month = {Aug},
  publisher = {American Physical Society},
  doi = {10.1103/PhysRevLett.101.080403},
  url = {https://link.aps.org/doi/10.1103/PhysRevLett.101.080403}
}

@article{
doi:10.1126/science.aan3211,
author = {Juan Yin  and Yuan Cao  and Yu-Huai Li  and Sheng-Kai Liao  and Liang Zhang  and Ji-Gang Ren  and Wen-Qi Cai  and Wei-Yue Liu  and Bo Li  and Hui Dai  and Guang-Bing Li  and Qi-Ming Lu  and Yun-Hong Gong  and Yu Xu  and Shuang-Lin Li  and Feng-Zhi Li  and Ya-Yun Yin  and Zi-Qing Jiang  and Ming Li  and Jian-Jun Jia  and Ge Ren  and Dong He  and Yi-Lin Zhou  and Xiao-Xiang Zhang  and Na Wang  and Xiang Chang  and Zhen-Cai Zhu  and Nai-Le Liu  and Yu-Ao Chen  and Chao-Yang Lu  and Rong Shu  and Cheng-Zhi Peng  and Jian-Yu Wang  and Jian-Wei Pan },
title = {Satellite-based entanglement distribution over 1200 kilometers},
journal = {Science},
volume = {356},
number = {6343},
pages = {1140-1144},
year = {2017},
doi = {10.1126/science.aan3211},
URL = {https://www.science.org/doi/abs/10.1126/science.aan3211},
eprint = {https://www.science.org/doi/pdf/10.1126/science.aan3211}}

@Article{Liao2017,
author={Liao, Sheng-Kai
and Cai, Wen-Qi
and Liu, Wei-Yue
and Zhang, Liang
and Li, Yang
and Ren, Ji-Gang
and Yin, Juan
and Shen, Qi
and Cao, Yuan
and Li, Zheng-Ping
and Li, Feng-Zhi
and Chen, Xia-Wei
and Sun, Li-Hua
and Jia, Jian-Jun
and Wu, Jin-Cai
and Jiang, Xiao-Jun
and Wang, Jian-Feng
and Huang, Yong-Mei
and Wang, Qiang
and Zhou, Yi-Lin
and Deng, Lei
and Xi, Tao
and Ma, Lu
and Hu, Tai
and Zhang, Qiang
and Chen, Yu-Ao
and Liu, Nai-Le
and Wang, Xiang-Bin
and Zhu, Zhen-Cai
and Lu, Chao-Yang
and Shu, Rong
and Peng, Cheng-Zhi
and Wang, Jian-Yu
and Pan, Jian-Wei},
title={Satellite-to-ground quantum key distribution},
journal={Nature},
year={2017},
month={Sep},
day={01},
volume={549},
number={7670},
pages={43-47},
issn={1476-4687},
doi={10.1038/nature23655},
url={https://doi.org/10.1038/nature23655}
}

@Article{Ren2017,
author={Ren, Ji-Gang
and Xu, Ping
and Yong, Hai-Lin
and Zhang, Liang
and Liao, Sheng-Kai
and Yin, Juan
and Liu, Wei-Yue
and Cai, Wen-Qi
and Yang, Meng
and Li, Li
and Yang, Kui-Xing
and Han, Xuan
and Yao, Yong-Qiang
and Li, Ji
and Wu, Hai-Yan
and Wan, Song
and Liu, Lei
and Liu, Ding-Quan
and Kuang, Yao-Wu
and He, Zhi-Ping
and Shang, Peng
and Guo, Cheng
and Zheng, Ru-Hua
and Tian, Kai
and Zhu, Zhen-Cai
and Liu, Nai-Le
and Lu, Chao-Yang
and Shu, Rong
and Chen, Yu-Ao
and Peng, Cheng-Zhi
and Wang, Jian-Yu
and Pan, Jian-Wei},
title={Ground-to-satellite quantum teleportation},
journal={Nature},
year={2017},
month={Sep},
day={01},
volume={549},
number={7670},
pages={70-73},
issn={1476-4687},
doi={10.1038/nature23675},
url={https://doi.org/10.1038/nature23675}
}

@article{PhysRevA.95.032306,
  title = {Entanglement swapping with independent sources over an optical-fiber network},
  author = {Sun, Qi-Chao and Mao, Ya-Li and Jiang, Yang-Fan and Zhao, Qi and Chen, Si-Jing and Zhang, Wei and Zhang, Wei-Jun and Jiang, Xiao and Chen, Teng-Yun and You, Li-Xing and Li, Li and Huang, Yi-Dong and Chen, Xian-Feng and Wang, Zhen and Ma, Xiongfeng and Zhang, Qiang and Pan, Jian-Wei},
  journal = {Phys. Rev. A},
  volume = {95},
  issue = {3},
  pages = {032306},
  numpages = {6},
  year = {2017},
  month = {Mar},
  publisher = {American Physical Society},
  doi = {10.1103/PhysRevA.95.032306},
  url = {https://link.aps.org/doi/10.1103/PhysRevA.95.032306}
}

@article{Sun:17,
author = {Qi-Chao Sun and Yang-Fan Jiang and Ya-Li Mao and Li-Xing You and Wei Zhang and Wei-Jun Zhang and Xiao Jiang and Teng-Yun Chen and Hao Li and Yi-Dong Huang and Xian-Feng Chen and Zhen Wang and Jingyun Fan and Qiang Zhang and Jian-Wei Pan},
journal = {Optica},
number = {10},
pages = {1214--1218},
publisher = {Optica Publishing Group},
title = {Entanglement swapping over 100\&\#x2009;\&\#x2009;km optical fiber with independent entangled photon-pair sources},
volume = {4},
month = {Oct},
year = {2017},
url = {https://opg.optica.org/optica/abstract.cfm?URI=optica-4-10-1214},
doi = {10.1364/OPTICA.4.001214},
}

@article{PhysRevA.77.022312,
  title = {Effects of spectral entanglement in polarization-entanglement swapping and type-I fusion gates},
  author = {Humble, Travis S. and Grice, Warren P.},
  journal = {Phys. Rev. A},
  volume = {77},
  issue = {2},
  pages = {022312},
  numpages = {9},
  year = {2008},
  month = {Feb},
  publisher = {American Physical Society},
  doi = {10.1103/PhysRevA.77.022312},
  url = {https://link.aps.org/doi/10.1103/PhysRevA.77.022312}
}

@article{PhysRevA.106.063711,
  title = {Generalized description of the spatio-temporal biphoton state in spontaneous parametric down-conversion},
  author = {Baghdasaryan, Baghdasar and Sevilla-Guti\'errez, Carlos and Steinlechner, Fabian and Fritzsche, Stephan},
  journal = {Phys. Rev. A},
  volume = {106},
  issue = {6},
  pages = {063711},
  numpages = {8},
  year = {2022},
  month = {Dec},
  publisher = {American Physical Society},
  doi = {10.1103/PhysRevA.106.063711},
  url = {https://link.aps.org/doi/10.1103/PhysRevA.106.063711}
}

@article{PhysRevLett.80.3891,
  title = {Experimental Entanglement Swapping: Entangling Photons That Never Interacted},
  author = {Pan, Jian-Wei and Bouwmeester, Dik and Weinfurter, Harald and Zeilinger, Anton},
  journal = {Phys. Rev. Lett.},
  volume = {80},
  issue = {18},
  pages = {3891--3894},
  numpages = {0},
  year = {1998},
  month = {May},
  publisher = {American Physical Society},
  doi = {10.1103/PhysRevLett.80.3891},
  url = {https://link.aps.org/doi/10.1103/PhysRevLett.80.3891}
}

@article{PhysRevLett.88.017903,
  title = {Experimental Nonlocality Proof of Quantum Teleportation and Entanglement Swapping},
  author = {Jennewein, Thomas and Weihs, Gregor and Pan, Jian-Wei and Zeilinger, Anton},
  journal = {Phys. Rev. Lett.},
  volume = {88},
  issue = {1},
  pages = {017903},
  numpages = {4},
  year = {2001},
  month = {Dec},
  publisher = {American Physical Society},
  doi = {10.1103/PhysRevLett.88.017903},
  url = {https://link.aps.org/doi/10.1103/PhysRevLett.88.017903}
}

@article{PhysRevLett.71.4287,
  title = {``Event-ready-detectors'' Bell experiment via entanglement swapping},
  author = {\ifmmode \dot{Z}\else \.{Z}\fi{}ukowski, M. and Zeilinger, A. and Horne, M. A. and Ekert, A. K.},
  journal = {Phys. Rev. Lett.},
  volume = {71},
  issue = {26},
  pages = {4287--4290},
  numpages = {0},
  year = {1993},
  month = {Dec},
  publisher = {American Physical Society},
  doi = {10.1103/PhysRevLett.71.4287},
  url = {https://link.aps.org/doi/10.1103/PhysRevLett.71.4287} 
}

@Article{Azuma2015,
author={Azuma, Koji
and Tamaki, Kiyoshi
and Lo, Hoi-Kwong},
title={All-photonic quantum repeaters},
journal={Nature Communications},
year={2015},
month={Apr},
day={15},
volume={6},
number={1},
pages={6787},
issn={2041-1723},
doi={10.1038/ncomms7787},
url={https://doi.org/10.1038/ncomms7787}
}

@article{PhysRevLett.133.233601,
  title = {Entangling Independent Particles by Path Identity},
  author = {Wang, Kai and Hou, Zhaohua and Qian, Kaiyi and Chen, Leizhen and Krenn, Mario and Zhu, Shining and Ma, Xiao-Song},
  journal = {Phys. Rev. Lett.},
  volume = {133},
  issue = {23},
  pages = {233601},
  numpages = {6},
  year = {2024},
  month = {Dec},
  publisher = {American Physical Society},
  doi = {10.1103/PhysRevLett.133.233601},
  url = {https://link.aps.org/doi/10.1103/PhysRevLett.133.233601}
}

@article{Tsujimoto:21,
author = {Yoshiaki Tsujimoto and Kentaro Wakui and Mikio Fujiwara and Masahide Sasaki and Masahiro Takeoka},
journal = {Opt. Express},
number = {23},
pages = {37150--37160},
publisher = {Optica Publishing Group},
title = {Ultra-fast Hong-Ou-Mandel interferometry via temporal filtering},
volume = {29},
month = {Nov},
year = {2021},
url = {https://opg.optica.org/oe/abstract.cfm?URI=oe-29-23-37150},
doi = {10.1364/OE.430502},
}

@article{Wakui:20,
author = {Kentaro Wakui and Yoshiaki Tsujimoto and Mikio Fujiwara and Isao Morohashi and Tadashi Kishimoto and Fumihiro China and Masahiro Yabuno and Shigehito Miki and Hirotaka Terai and Masahide Sasaki and Masahiro Takeoka},
journal = {Opt. Express},
number = {15},
pages = {22399--22411},
publisher = {Optica Publishing Group},
title = {Ultra-high-rate nonclassical light source with 50\&\#x2009;GHz-repetition-rate mode-locked pump pulses and multiplexed single-photon detectors},
volume = {28},
month = {Jul},
year = {2020},
url = {https://opg.optica.org/oe/abstract.cfm?URI=oe-28-15-22399},
doi = {10.1364/OE.397030},
}

@Article{Bedington2017,
author={Bedington, Robert
and Arrazola, Juan Miguel
and Ling, Alexander},
title={Progress in satellite quantum key distribution},
journal={npj Quantum Information},
year={2017},
month={Aug},
day={09},
volume={3},
number={1},
pages={30},
issn={2056-6387},
doi={10.1038/s41534-017-0031-5},
url={https://doi.org/10.1038/s41534-017-0031-5}
}

@article{PhysRevApplied.19.054082,
  title = {Clock Synchronization with Correlated Photons},
  author = {Spiess, Christopher and T\"opfer, Sebastian and Sharma, Sakshi and Kr\ifmmode \check{z}\else \v{z}\fi{}i\ifmmode \check{c}\else \v{c}\fi{}, Andrej and Cabrejo-Ponce, Meritxell and Chandrashekara, Uday and D\"oll, Nico Lennart and Riel\"ander, Daniel and Steinlechner, Fabian},
  journal = {Phys. Rev. Appl.},
  volume = {19},
  issue = {5},
  pages = {054082},
  numpages = {22},
  year = {2023},
  month = {May},
  publisher = {American Physical Society},
  doi = {10.1103/PhysRevApplied.19.054082},
  url = {https://link.aps.org/doi/10.1103/PhysRevApplied.19.054082}
}

@article{Spiess_2024,
doi = {10.1088/2058-9565/ad0ce0},
url = {https://doi.org/10.1088/2058-9565/ad0ce0},
year = {2023},
month = {dec},
publisher = {IOP Publishing},
volume = {9},
number = {1},
pages = {015019},
author = {Spiess, Christopher and Steinlechner, Fabian},
title = {Clock synchronization with pulsed single photon sources},
journal = {Quantum Science and Technology}
}

@article{Sun2016,
  title={Quantum teleportation with independent sources and prior entanglement distribution over a network},
  author={Sun, Qi-Cai and Mao, Ya-Li and Chen, Si-Jing and Zhang, Wei and Yang, Yang-Fan and Mussot, Arnaud and Kudlinski, Alexandre and Kiefer, J and \dots and Pan, Jian-Wei},
  journal={Nature Photonics},
  volume={10},
  number={10},
  pages={671--675},
  year={2016},
  publisher={Nature Publishing Group}
}

@Article{Qian2023,
author={Qian, Kaiyi
and Wang, Kai
and Chen, Leizhen
and Hou, Zhaohua
and Krenn, Mario
and Zhu, Shining
and Ma, Xiao-song},
title={Multiphoton non-local quantum interference controlled by an undetected photon},
journal={Nature Communications},
year={2023},
month={Mar},
day={17},
volume={14},
number={1},
pages={1480},
issn={2041-1723},
doi={10.1038/s41467-023-37228-y},
url={https://doi.org/10.1038/s41467-023-37228-y}
}

@Article{Jin2015,
  title     = "Highly efficient entanglement swapping and teleportation at
               telecom wavelength",
  author    = "Jin, Rui-Bo and Takeoka, Masahiro and Takagi, Utako and Shimizu,
               Ryosuke and Sasaki, Masahide",
  journal   = "Sci. Rep.",
  publisher = "Springer Science and Business Media LLC",
  volume    =  5,
  number    =  1,
  pages     = "9333",
  month     =  mar,
  year      =  2015,
  copyright = "https://creativecommons.org/licenses/by/4.0",
  language  = "en"
}

@Article{Keller2003,
  title     = "Recent developments in compact ultrafast lasers",
  author    = "Keller, Ursula",
  journal   = "Nature",
  publisher = "Springer Science and Business Media LLC",
  volume    =  424,
  number    =  6950,
  pages     = "831--838",
  month     =  aug,
  year      =  2003,
  language  = "en"
}

@inproceedings{AndrejCode,
author = {Andrej Krzic and Daniel Heinig and Matthias Goy and Fabian Steinlechner},
title = {{Dual-downlink quantum key distribution with entangled photons: prospects for daylight operation}},
volume = {12777},
booktitle = {International Conference on Space Optics — ICSO 2022},
publisher = {SPIE},
year = {2023},
doi = {10.1117/12.2689971},
URL = {https://doi.org/10.1117/12.2689971}
}

@Article{Zhang2017,
author={Zhang, Yingwen
and Agnew, Megan
and Roger, Thomas
and Roux, Filippus S.
and Konrad, Thomas
and Faccio, Daniele
and Leach, Jonathan
and Forbes, Andrew},
title={Simultaneous entanglement swapping of multiple orbital angular momentum states of light},
journal={Nature Communications},
year={2017},
month={Sep},
day={21},
volume={8},
number={1},
pages={632},
issn={2041-1723},
doi={10.1038/s41467-017-00706-1},
url={https://doi.org/10.1038/s41467-017-00706-1}
}

@Article{Halder2007,
author={Halder, Matth{\"a}us
and Beveratos, Alexios
and Gisin, Nicolas
and Scarani, Valerio
and Simon, Christoph
and Zbinden, Hugo},
title={Entangling independent photons by time measurement},
journal={Nature Physics},
year={2007},
month={Oct},
day={01},
volume={3},
number={10},
pages={692-695},
issn={1745-2481},
doi={10.1038/nphys700},
url={https://doi.org/10.1038/nphys700}
}

@article{PhysRevLett.94.083601,
  title = {Two-Photon Coincident-Frequency Entanglement via Extended Phase Matching},
  author = {Kuzucu, Onur and Fiorentino, Marco and Albota, Marius A. and Wong, Franco N. C. and K\"artner, Franz X.},
  journal = {Phys. Rev. Lett.},
  volume = {94},
  issue = {8},
  pages = {083601},
  numpages = {4},
  year = {2005},
  month = {Mar},
  publisher = {American Physical Society},
  doi = {10.1103/PhysRevLett.94.083601},
  url = {https://link.aps.org/doi/10.1103/PhysRevLett.94.083601}
}

@article{Halder_2008,
doi = {10.1088/1367-2630/10/2/023027},
url = {https://dx.doi.org/10.1088/1367-2630/10/2/023027},
year = {2008},
month = {feb},
publisher = {},
volume = {10},
number = {2},
pages = {023027},
author = {Halder, Matthäus and Beveratos, Alexios and Thew, Robert T and Jorel, Corentin and Zbinden, Hugo and Gisin, Nicolas},
title = {High coherence photon pair source for quantum communication},
journal = {New Journal of Physics}
}

@article{PhysRevA.82.043826,
  title = {Heralding single photons without spectral factorability},
  author = {Huang, Yu-Ping and Altepeter, Joseph B. and Kumar, Prem},
  journal = {Phys. Rev. A},
  volume = {82},
  issue = {4},
  pages = {043826},
  numpages = {6},
  year = {2010},
  month = {Oct},
  publisher = {American Physical Society},
  doi = {10.1103/PhysRevA.82.043826},
  url = {https://link.aps.org/doi/10.1103/PhysRevA.82.043826}
}

@article{PRXQuantum.6.020304,
  title = {Tailoring Fusion-Based Photonic Quantum Computing Schemes to Quantum Emitters},
  author = {Chan, Ming Lai and Bell, Thomas J. and Pettersson, Love A. and Chen, Susan X. and Yard, Patrick and S\o{}rensen, Anders S. and Paesani, Stefano},
  journal = {PRX Quantum},
  volume = {6},
  issue = {2},
  pages = {020304},
  numpages = {18},
  year = {2025},
  month = {Apr},
  publisher = {American Physical Society},
  doi = {10.1103/PRXQuantum.6.020304},
  url = {https://link.aps.org/doi/10.1103/PRXQuantum.6.020304}
}

@article{Moschandreou:18,
author = {Eleftherios Moschandreou and Jeffrey I. Garcia and Brian J. Rollick and Bing Qi and Raphael Pooser and George Siopsis},
journal = {J. Lightwave Technol.},
number = {17},
pages = {3752--3759},
publisher = {Optica Publishing Group},
title = {Experimental Study of Hong--Ou--Mandel Interference Using Independent Phase Randomized Weak Coherent States},
volume = {36},
month = {Sep},
year = {2018},
url = {https://opg.optica.org/jlt/abstract.cfm?URI=jlt-36-17-3752},
}

@article{Samara:19,
author = {Farid Samara and Anthony Martin and Claire Autebert and Maxim Karpov and Tobias J. Kippenberg and Hugo Zbinden and Rob Thew},
journal = {Opt. Express},
number = {14},
pages = {19309--19318},
publisher = {Optica Publishing Group},
title = {High-rate photon pairs and sequential Time-Bin entanglement with Si3N4 microring resonators},
volume = {27},
month = {Jul},
year = {2019},
url = {https://opg.optica.org/oe/abstract.cfm?URI=oe-27-14-19309},
doi = {10.1364/OE.27.019309},
}

@article{Natarajan_2012,
doi = {10.1088/0953-2048/25/6/063001},
url = {https://doi.org/10.1088/0953-2048/25/6/063001},
year = {2012},
month = {apr},
publisher = {IOP Publishing},
volume = {25},
number = {6},
pages = {063001},
author = {Natarajan, Chandra M and Tanner, Michael G and Hadfield, Robert H},
title = {Superconducting nanowire single-photon detectors: physics and applications},
journal = {Superconductor Science and Technology}
}

@ARTICLE{618322,
  author={Erdogan, T.},
  journal={Journal of Lightwave Technology}, 
  title={Fiber grating spectra}, 
  year={1997},
  volume={15},
  number={8},
  pages={1277-1294},
  doi={10.1109/50.618322}}

@article{Kraemer2025,
  author       = {Krämer, R. G. and Schmittner, C. and Ullsperger, T. and Siems, M. P. and Döpfner, S. L. and Schwartz, G. R. and Richter, D. and Nolte, S.},
  title        = {Shaping the spectral response of ultrashort pulse phase mask written fiber Bragg gratings},
  journal      = {Optics Express},
  volume       = {33},
  number       = {13},
  pages        = {28152–28163},
  month        = jun,
  year         = {2025},
  doi          = {10.1364/OE.558398},
}

@article{PhysRevA.78.032112,
  title = {More efficient Bell inequalities for Werner states},
  author = {V\'ertesi, T.},
  journal = {Phys. Rev. A},
  volume = {78},
  issue = {3},
  pages = {032112},
  numpages = {6},
  year = {2008},
  month = {Sep},
  publisher = {American Physical Society},
  doi = {10.1103/PhysRevA.78.032112},
  url = {https://link.aps.org/doi/10.1103/PhysRevA.78.032112}
}

@article{Bouchard_2021,
doi = {10.1088/1361-6633/abcd7a},
url = {https://dx.doi.org/10.1088/1361-6633/abcd7a},
year = {2020},
month = {dec},
publisher = {IOP Publishing},
volume = {84},
number = {1},
pages = {012402},
author = {Frédéric Bouchard and Alicia Sit and Yingwen Zhang and Robert Fickler and Filippo M Miatto and Yuan Yao and Fabio Sciarrino and Ebrahim Karimi},
title = {Two-photon interference: the Hong–Ou–Mandel effect},
journal = {Reports on Progress in Physics}
}

@article{PhysRevLett.59.2044,
  title = {Measurement of subpicosecond time intervals between two photons by interference},
  author = {Hong, C. K. and Ou, Z. Y. and Mandel, L.},
  journal = {Phys. Rev. Lett.},
  volume = {59},
  issue = {18},
  pages = {2044--2046},
  numpages = {0},
  year = {1987},
  month = {Nov},
  publisher = {American Physical Society},
  doi = {10.1103/PhysRevLett.59.2044},
  url = {https://link.aps.org/doi/10.1103/PhysRevLett.59.2044}
}

@article{Tsujimoto:17,
author = {Yoshiaki Tsujimoto and Yukihiro Sugiura and Motoki Tanaka and Rikizo Ikuta and Shigehito Miki and Taro Yamashita and Hirotaka Terai and Mikio Fujiwara and Takashi Yamamoto and Masato Koashi and Masahide Sasaki and Nobuyuki Imoto},
journal = {Opt. Express},
number = {11},
pages = {12069--12080},
publisher = {Optica Publishing Group},
title = {High visibility Hong-Ou-Mandel interference via a time-resolved coincidence measurement},
volume = {25},
month = {May},
year = {2017},
url = {https://opg.optica.org/oe/abstract.cfm?URI=oe-25-11-12069},
doi = {10.1364/OE.25.012069},
}

@article{PhysRevA.109.023534,
  title = {Spectral properties of transverse Laguerre-Gauss modes in parametric down-conversion},
  author = {Sevilla-Guti\'errez, Carlos and Kaipalath, Varun Raj and Baghdasaryan, Baghdasar and Gr\"afe, Markus and Fritzsche, Stephan and Steinlechner, Fabian},
  journal = {Phys. Rev. A},
  volume = {109},
  issue = {2},
  pages = {023534},
  numpages = {11},
  year = {2024},
  month = {Feb},
  publisher = {American Physical Society},
  doi = {10.1103/PhysRevA.109.023534},
  url = {https://link.aps.org/doi/10.1103/PhysRevA.109.023534}
}

@article{TAKESUE2010276,
title = {Effects of multiple pairs on visibility measurements of entangled photons generated by spontaneous parametric processes},
journal = {Optics Communications},
volume = {283},
number = {2},
pages = {276-287},
year = {2010},
issn = {0030-4018},
doi = {https://doi.org/10.1016/j.optcom.2009.10.008},
url = {https://www.sciencedirect.com/science/article/pii/S0030401809009651},
author = {Hiroki Takesue and Kaoru Shimizu}
}

@incollection{LEGERO2006253,
title = {Characterization of Single Photons Using Two-Photon Interference},
editor = {G. Rempe and M.O. Scully},
series = {Advances In Atomic, Molecular, and Optical Physics},
publisher = {Academic Press},
volume = {53},
pages = {253-289},
year = {2006},
issn = {1049-250X},
doi = {https://doi.org/10.1016/S1049-250X(06)53009-5},
url = {https://www.sciencedirect.com/science/article/pii/S1049250X06530095},
author = {T. Legero and T. Wilk and A. Kuhn and G. Rempe}
}

@article{PhysRevA.98.013833,
  title = {Quantum description of timing jitter for single-photon ON-OFF detectors},
  author = {Gouzien, \'Elie and Fedrici, Bruno and Zavatta, Alessandro and Tanzilli, S\'ebastien and D'Auria, Virginia},
  journal = {Phys. Rev. A},
  volume = {98},
  issue = {1},
  pages = {013833},
  numpages = {5},
  year = {2018},
  month = {Jul},
  publisher = {American Physical Society},
  doi = {10.1103/PhysRevA.98.013833},
  url = {https://link.aps.org/doi/10.1103/PhysRevA.98.013833}
}

@article{Samara_2021,
doi = {10.1088/2058-9565/abf599},
url = {https://dx.doi.org/10.1088/2058-9565/abf599},
year = {2021},
month = {sep},
publisher = {IOP Publishing},
volume = {6},
number = {4},
pages = {045024},
author = {Farid Samara and Nicolas Maring and Anthony Martin and Arslan S Raja and Tobias J Kippenberg and Hugo Zbinden and Rob Thew},
title = {Entanglement swapping between independent and asynchronous integrated photon-pair sources},
journal = {Quantum Science and Technology}
}

@Article{Tsujimoto2018,
author={Tsujimoto, Yoshiaki
and Tanaka, Motoki
and Iwasaki, Nobuo
and Ikuta, Rikizo
and Miki, Shigehito
and Yamashita, Taro
and Terai, Hirotaka
and Yamamoto, Takashi
and Koashi, Masato
and Imoto, Nobuyuki},
title={High-fidelity entanglement swapping and generation of three-qubit GHZ state using asynchronous telecom photon pair sources},
journal={Scientific Reports},
year={2018},
month={Jan},
day={23},
volume={8},
number={1},
pages={1446},
issn={2045-2322},
doi={10.1038/s41598-018-19738-8},
url={https://doi.org/10.1038/s41598-018-19738-8} 
}

@article{PhysRevLett.93.070503,
  title = {Quantum Beat of Two Single Photons},
  author = {Legero, Thomas and Wilk, Tatjana and Hennrich, Markus and Rempe, Gerhard and Kuhn, Axel},
  journal = {Phys. Rev. Lett.},
  volume = {93},
  issue = {7},
  pages = {070503},
  numpages = {4},
  year = {2004},
  month = {Aug},
  publisher = {American Physical Society},
  doi = {10.1103/PhysRevLett.93.070503},
  url = {https://link.aps.org/doi/10.1103/PhysRevLett.93.070503}
}

@Article{Legero2003,
author={Legero, T.
and Wilk, T.
and Kuhn, A.
and Rempe, G.},
title={Time-resolved two-photon quantum interference},
journal={Applied Physics B},
year={2003},
month={Dec},
day={01},
volume={77},
number={8},
pages={797-802},
issn={1432-0649},
doi={10.1007/s00340-003-1337-x},
url={https://doi.org/10.1007/s00340-003-1337-x}
}

\end{document}